\documentclass[aps,prb,twocolumn]{revtex4}
\usepackage{bm,color,amsmath,amssymb,mathrsfs,latexsym,graphicx,psfrag}

\newcommand{\im}{\mathrm{Im}}

\newcommand{\bk}{{\mathbf k}}
\newcommand{\bq}{{\mathbf q}}
\newcommand{\br}{{\mathbf{r}}}

\newcommand{\be}{\begin{equation}}
\newcommand{\ee}{\end{equation}}

\newcommand{\bK}{\mathbf{K}}

\def\be{\begin{equation}}
\def\ee{\end{equation}}
\def\bea{\begin{eqnarray}}
\def\eea{\end{eqnarray}}

\begin{document}
\title{Topological crystalline superconductors with linearly and projectively represented $C_{n}$ symmetry}
\author{Chen Fang}%
\affiliation{Beijing National Laboratory for Condensed Matter Physics and Institute of Physics, Chinese Academy of Sciences, Beijing 100190, China}%

\author{B. Andrei Bernevig}%
\affiliation{Department of Physics, Princeton University, Princeton NJ 08544}

\author{Matthew J. Gilbert}%
\affiliation{Micro and Nanotechnology Laboratory, University of Illinois, Urbana IL 61801}
\affiliation{Department of Electrical and Computer Engineering, University of Illinois, Urbana IL 61801}
\affiliation{Department of Electronics Engineering, University of Rome "Tor Vergata", Rome, Italy 00133}
\date{\today}

\begin{abstract}
We study superconductors with $n$-fold rotational invariance both in the presence and in the absence of spin-orbit interactions. More specifically, we classify the non-interacting Hamiltonians by defining a series of $Z$-numbers for the Bogoliubov-de Gennes (BdG) symmetry classes of the Altland-Zimbauer classification of random matrices in $1$D, $2$D, and $3$D in the presence of discrete rotational invariance. Our analysis emphasizes the important role played by the angular momentum of the Cooper pairs in the system: for pairings of nonzero angular momentum, the rotation symmetry may be represented projectively, and a projective representation of rotation symmetry may have anomalous properties, including the anti-commutation with the time-reversal symmetry. In 1D and 3D, we show how an $n$-fold axis enhances the topological classification and give additional topological numbers; in 2D, we establish a relation between the Chern number (in class D and CI) and the eigenvalues of rotation symmetry at high-symmetry points. For each nontrivial class in 3D, we write down a minimal effective theory for the surface Majorana states.
\end{abstract}

\maketitle

\section{Introduction}
Condensed matter physics has, in recent years, been partly focused on the search for new materials that harbor topological states. A topological state is a gapped many-body state that cannot be adiabatically connected to the atomic limit while preserving a certain symmetry group, and yet cannot be associated with any local order parameter. In place of order parameters, topological numbers, a global quantity contributed to by all the electrons in system, distinguish a topological state from a trivial one. Initial efforts have focused on the elucidation of topological states whose existence and global properties are stabilized by the presence of time-reversal symmetry (TRS). This search has led to the examination of a plethora of 2D\cite{kane2005A,kane2005B,bernevig2006a,bernevig2006c,koenig2007,roth2009,liu2008,Knez2011,Knez2012,Qian2014} and 3D non-interacting band insulating systems\cite{fu2007b,Qi2008,hsieh2009a,Hsieh2009,hsieh2009b,Chen:2009vn,xia2009,roushan2009,Alpichshev2010,chen2010,liu2010,Seo2010,Checkelsky2011,Okada2011,Chang2013} under a wide variety of experimental conditions seeking to explore the fundamental spin and charge behavior of TRS topological band insulators. Yet the underlying principles of symmetry preservation required for the stability of a topological phase within a given host material are quite general in nature thereby making the TRS class of topological non-interacting band insulators but one in a long list of candidate topological materials. Therefore, while we understand a great deal about the physical nature of TRS topological band insulators, we are at but the beginning in the search for topological materials\cite{Senthil2015,Ando2015,Chiu2015}.

From theoretical point of view, the discussion of topological materials beyond those that preserve TRS took a significant step forward when the Heusler class of materials were predicted as topological candidates\cite{Chadov2010,Lin:2010,Hirschberger2016,ShekharFelser2016}. While the specific focus of the original work\cite{Chadov2010,Lin:2010} had been to explore the existence of additional TRS topological band insulator candidates in the Heusler class of materials, the Heusler compounds exhibit an extremely wide range of physical phenomena such as ferromagnetism\cite{Canfield1991}, which expressly breaks TRS, and heavy fermion behavior\cite{Fisk1991}. This work provided early indications that multiple symmetries can be present in materials and establish different topological orders. Fu\cite{Fu2011} was the first to provide an explicit proof that, in 2D and 3D band insulating systems, the presence of rotational symmetry in the underlying lattice structure, namely $C_4$, together with time-reversal symmetry give rise to a new $Z_2$ classification for such insulators even in the absence of spin-orbit coupling. Such materials which have topological states whose existence is guaranteed by the presence of underlying crystalline symmetry are commonly referred to as topological crystalline insulators (TCI). From this early work, the search for topological materials beyond those with TRS has since been expanded. Predictions of other types that can be generally classified as topological crystalline systems have emerged, including that of inversion symmetric topological materials\cite{hughes2010inv,turner2010,Turner:2012}, and of rotationally invariant topological band insulators\cite{Fang2012}. Topological crystalline insulators protected by mirror reflection symmetry\cite{Hsieh2012,Liu2013,Liu2013nm,Fang2013,Fang2013a,Fang2012a} and glide reflection symmetries\cite{Liu2014,Fang2015,Shiosaki2015,Lu2016,Wang2016,Alexandradinata2016} have been theoretically studied and some have been confirmed in experiments\cite{Dziawa2012,Tanaka2012,Xu2012,Ma2016}. Most recently, predictions of topological semimetals whose band crossings are protected by rotational invariance\cite{Fang2012b,XuG2011,Wang2012} have also seen experimental confirmation in Na$_{3}$Bi\cite{Liu2013sm} and in Cd$_{3}$As$_{2}$\cite{Neupane2013,Borisenko2013}.

One can further consider the case where particle number is no longer conserved - that is, the case of superconductors, represented by a Bogoliubov-de Gennes (BdG) Hamiltonian. Ref.[\onlinecite{schnyder2008,Ryu2010,schnyder2011}] studied the topological classifications of fully gapped superconductors in all dimensions both in the presence and absence of time-reversal symmetry and spin-rotation symmetry. This initial work has been followed by further research into topological phases in superconductors containing various symmetries such as: TRS superconductors\cite{qi2009trstsc,Qi2010,Fu2010}, reflection symmetric superconductors\cite{Zhang2013,Chiu2013}, non-centrosymmetric superconductors\cite{schnyder2011}, topological superfluid $^{3}$He-B\cite{volovik1988,qi2009trstsc}, and Weyl superconductors\cite{Cho2012,shivamoggi2013}. Nevertheless, the superconducting phase of materials with general rotational invariance has remained relatively unstudied (yet see Ref.[\onlinecite{Shiozaki2014}] for a discussion of superconductors with twofold symmetries). Rotational symmetries are important in understanding the superconducting behavior demonstrated in the Heusler alloys LaBiPt\cite{Goll2008} and, most recently, in YPtBi\cite{Butch2011}. Hence there is a need for theoretical elucidation of the possible corresponding topological nature of such superconducting systems.

In this work we study topological superconductors that possess $C_{n}$ rotational symmetry with and without spin-orbital interactions, and ask if the presence of $C_{n}$ symmetry can stabilize additional topological crystalline superconductors. We also explore how the $C_n$ symmetry places constraints on the invariants of the original BdG classes. We ask this question broadly by considering the behavior of each of the $4$ distinct Bogoliubov- de Gennes (BdG) symmetry classes, namely class C, D, CI and DIII, of random matrices from the Altland and Zimbauer (AZ)\cite{altland1997} in $1$D, $2$D, and $3$D. The answer to such a question has direct relevance to a wide range of unconventional superconducting materials such as cuprates and iron-based superconductors. In Section \ref{sec:prelim}, we discuss the necessary general background to understand the subsequent analysis. In Section \ref{sec:1d}, we perform a complete classification of $1$D superconductors with rotational symmetries. In the classification, we find it important to distinguish the case where the $C_n$ symmetry is represented projectively from the case where it is linearly represented: in the former case, time-reversal symmetry \emph{anti-commutes} with the rotation symmetry when acting on a Bogoliubov quasiparticle. Which case appears depends on the total angular momentum of the Cooper pair. In Section \ref{sec:2d}, we consider Chern superconductors in $2$D with rotational symmetries, deriving explicit relations between the Chern number and the rotation eigenvalues of occupied bands at high-symmetry points. In Section \ref{sec:3d}, we apply our knowledge of $1$D superconductors to classify both high-symmetry and generic lines within the Brillouin zone (BZ) of $3$D superconductors by determining not only the bulk invariants but also the effective surface theory. In Section \ref{sec:conclusion}, we summarize our results and their implications.

\section{Preliminaries}
\label{sec:prelim}
\subsection{Enhancement of the AZ Classification by Local Unitary Symmetries}

The AZ classification of single particle Hamiltonians\cite{altland1997} is based on the transformation properties of the single particle Hamiltonian under two local symmetries, namely, particle-hole symmetry (PHS, or $P$) and time-reversal symmetry (TRS, or $T$), and their composition which is called chiral symmetry. A non-interacting Hamiltonian here refers to a Hamiltonian that only has quadratic couplings among the creation and annihilation operators (in the second quantized form), and can be put into a matrix, $H$, in the orbital basis (or Nambu basis when charge is not conserved) in the first quantized form. The two local antiunitary symmetries are, given a basis, represented by $KT$ and $KP$, where $T,P$ are unitary matrices and $K$ complex conjugation. A Hamiltonian $H$ is said to have $T$ if and only if $[KT,H]=0$ and have $P$ if and only if $\{KP,H\}=0$. We remark that these formulas apply in the first quantized form, where all operators are represented by matrices, while in the second quantized form, where operators are expanded in terms of fermion annihilation and creation operators, both P and T should commute with H, and P becomes a \emph{unitary} operator. Hereafter, we use hatted symbols for second quantized operators, and non-hatted ones for their first quantized forms.When both $P$ and $T$ are present, we can define $S=KP*KT=P^\ast*{T}$ such that $\{S,H\}=0$, and we say the system has chiral symmetry $S$. However, there are cases where $S$ is a symmetry, i.e., $\{S,H\}=0$, while neither $KP$ nor $KT$ is a symmetry.

Now we consider a local unitary symmetry added to the system, represented by some unitary matrix $L$, which generically satisfies
\bea
[{L},{H}]=0.
\eea
Therefore the Hamiltonian can be block-diagonalized into sectors spanned by eigenvectors of ${L}$, namely,
\bea
{H}={H}_{r_1}\oplus{H}_{r_2}\oplus...+{H}_{r_l},
\eea
where $s$ is the number of eigenvalues of ${L}$, and $r_{1,...,l}$ the eigenvalues; in sector $r_i$, the basis vectors are eigenstates of $L$ with eigenvalue $r_i$. For each sector, we can classify each ${H}_{r_i}$ according to its transformation under time-reversal, particle-hole and chiral symmetries. Physically, time-reversal and particle-hole symmetries commute with all spatial symmetries, and let us assume that $[{KT},{L}]=[{KP},{L}]=0$. However, one is reminded that generally $[{KT},{H}_{r_i}]\neq0$ and $[{KP},{H}_{r_i}]\neq0$. This is because if $r_i\notin{Real}$, $KP$ or $KT$ sends the state to another sector with eigenvalue $r^\ast_i$. However the chiral symmetry, represented by a unitary matrix, preserves the block structure of $H$.

Based on this discussion, we come to a simple conclusion: for any $r\in{Real}$, the Hamiltonian ${H}_{r}$ inherits the particle-hole, time-reversal and chiral symmetries the system may have, thus having the same topological classification as that found Ref.[\onlinecite{schnyder2008,Ryu2010,schnyder2011}]. On the other hand, if $r\notin{Real}$, ${H}_{r}$ only inherits the chiral symmetry of the system, should the system possess chiral symmetry, while the time-reversal and particle-hole symmetries relate ${H}_{r}$ to ${H}_{r^\ast}$. Due to this relation, the topological invariants for ${H}_r$ and ${H}_{r^\ast}$ will be shown to be either the same or opposite, depending on $S$ and the symmetry class of the Hamiltonian. Therefore, a local unitary symmetry in general enhances the topological classifications, as now the full system is labeled by all the quantum numbers from each sector (if $r\in{Real}$) and each pair of sectors (if $r\notin{Real}$), rather than the single $Z$ or $Z_2$ number for the entire Hamiltonian $H$.

A point group, or specifically rotational, symmetry is in general \emph{not} a local symmetry, as it changes the position of an electron. The only exceptions are mirror reflection in a 2D system when then mirror plane and the system are coplanar, and the rotation in a 1D system when the rotation axis coincides with the system. Nonetheless, for single particle Hamiltonians with translational symmetry, there are always some sub-manifolds in the $\bk$-space that are invariant under a point group symmetry. For example, in a 3D simple cubic lattice, the tight-binding Hamiltonian $H(k_x,k_y,k_z)$ in $\bk$-space is invariant under mirror reflection $M_{xy}:z\rightarrow{-z}$ when $k_z=0$ or $k_z=\pi$; and it is invariant under the fourfold rotation about the $z$-axis $C_4:(x,y,z)\rightarrow(-y,x,z)$, if $(k_x,k_y)=(0,0)$ or $(\pi,\pi)$, where the lattice constant is taken to be unity, $a\equiv1$. For an invariant sub-manifold, the point group symmetry becomes a local symmetry, and hence, we can use the general scheme described above, i.e., classifying the sectors labeled by the eigenvalues of $L$, to find the classification and the corresponding invariants of the Hamiltonian on the sub-manifold. The set of quantum numbers defined on all these invariant sub-manifolds characterize a general topological crystalline insulator or superconductor. It is this method that we will be using to classify superconductors with rotation symmetry within the context of this work. We should note, however, that this characterization is incomplete and there exist interesting exceptions\cite{Fu2011,Alexandradinata2014,Teo2013tcs}. For example, in Ref.[\onlinecite{Fu2011,Alexandradinata2014}], the authors show that 3D spinless systems host topological numbers that are protected by rotation symmetry but yet cannot be defined on any of the high-symmetry lines.

\subsection{Spinless Vs. Spinful Systems}

Having clarified the role of rotational symmetries in general Hamiltonians, it is important to mark the difference between spinless and spinful systems. As superconductors are, naturally, electronic systems and their constituent electrons are elementary particles with spin one-half, this distinction does not refer to the spin state of an electron. For our purposes, the term `spinless' simply refers to the unbroken SU(2) spin rotation symmetry, while the term `spinful' indicates its absence. When the spin-orbit interaction is ignored, an SU(2) invariant Hamiltonian can be block-diagonalized into two parts, namely those of spin up and the spin down parts, while both the TRS and PHS can be composed with a spin-rotation about a specific directional axis (say, $y$-axis) so as to not change the spin state, thus acting within each block. This signifies that the symmetries of the whole system completely pass to each of the respective spin sectors, and it is therefore sufficient to study any one of the two independent `spinless' Hamiltonians to understand the properties of the system as a whole.

However, this is not quite the complete picture of the proper physics. One needs to take caution in that since PHS and TRS are now combined with a spin rotation, their squares change sign, resulting in ${KT}^2=-{KP}^2=1$. This becomes a vital distinction and is needed when we discuss rotational symmetries. A full rotation of an electron gives a $-1$ factor to the wavefunctions due to the inherent $\pi$ Berry's phase. This points to the fact that for a general electronic system, we have ${C}_n^n=-1$ where ${C}_n$ is the rotation operator of an $n$-fold rotation. When SU(2) is present, the rotation symmetry can be redefined as a rotation of both spin and space followed by a spin rotation in the opposite direction. In such a case, we have ${C}_n^n=1$. Physically this means that in the absence of SOC, a rotation symmetry only operates on the spatial degrees of freedom, leaving the spin part unchanged. Within the context of operators, the terms `spinless' and `spinful' indicate ${C}_n^n={KT}^2=-{KP}^2=1$ and $-1$, respectively.

\subsection{Rotational Symmetry of a BdG Hamiltonian}

We are interested in discussing the properties of mean-field superconducting systems, and hence we must understand how the previously discussed rotational symmetries manifest themselves in a Bogoliubov-de Gennes (BdG) Hamiltonian. The second-quantized form of a BdG Hamiltonian reads:
\bea
\hat{H}=\hat{H}_0+\hat{\Delta}+\hat{\Delta}^\dag,
\eea
where
\bea
\hat{H}_0&=&H_{ab}c^\dag_ac_b+h.c.,\\
\nonumber
\hat{\Delta}&=&\Delta_{ab}c_ac_b,
\eea
where $a,b$ are composed indices labeling the site, orbital and spin in a lattice system.
In an $n$-fold rotation invariant system, we require that
\bea
\label{eq:rotcomnorm}
[\hat{C}_n,\hat{H}_0]=0,
\eea
and
\bea\label{eq:Delta}
\hat{C}_n\hat\Delta\hat{C}_n^{-1}=e^{i\theta}\hat\Delta.
\eea
Using the fact that $\hat{C}_n^n=\pm1$ for spinless and spinful systems respectively in conjunction with Eq.(\ref{eq:Delta}), we find $e^{in\theta}=1$ ,or $\theta=2m\pi/n$ where $m=0,...,n-1$. Consider a gauge transform $\hat{U}=\exp(im\pi/n\hat{Q})$, where $\hat{Q}$ is the total electric charge operator, such that
\bea
\hat{U}_mc_a\hat{U}^{-1}_m=e^{-im\pi/n}c_a,
\eea
and if we combine $\hat{U}_m$ with $\hat{C}_n$, from Eq. \ref{eq:rotcomnorm}, we have
\bea
[\hat{C}_n\hat{U}_m,\hat{H}]=0.
\eea
Therefore, we may define $\hat{C}_{n,m}\equiv\hat{C}_n\hat{U}_m$ as a symmetry of the system. In a translation invariant system, we have
\bea\label{eq:CnGInvariance}
C_{n,m}H(\bk)C_{n,m}^{-1}=H(C_n\bk),
\eea
where $C_{n,m}$ is the first quantized matrix representation of $\hat{C}_{n,m}$ in the Nambu basis. The significance of Eq.(\ref{eq:CnGInvariance}) lies in that the spectrum of $H(\bk)$ remains rotationally invariant even when $m\neq0$. Physically, $m\neq0$ indicates that the Cooper pair has total angular momentum $m\hbar$ along the rotation axis (modulo $n$), because the pair gains a phase of $e^{i2m\pi/n}$ after a rotation through $2\pi/n$. Nevertheless, this change of phase is not reflected in the quasiparticle spectrum, which is still $n$-fold symmetric, due to Eq.(\ref{eq:CnGInvariance}). This is because the phase can be compensated by a U(1) transform which leaves $\hat{H}_0$ invariant. Additionally, we should also point out that
\bea\label{eq:CnmandCn}
{C}_{n,m}^n=(-1)^m\hat{C}_n^n=(-1)^{m+F},
\eea
where $F=0,1$ for spinless and spinful fermions, respectively. Therefore, when $m\in{odd}$, ${C}_{n,m}$ is like a spinless (spinful) rotation in a spinful (spinless) system. Mathematically, when $m\in{odd}$, Eq.(\ref{eq:CnmandCn}) indicates that $C_{n,m}$ and $KP$ form a projective representation of the group generated by $C_n$ and $P$, a fact that we rigorously prove in Appendix \ref{app:PC}.

\subsection{Time-reversal symmetry and projective representation of the group generated by $C_n$ and $T$}
\label{sec:project}
In this paper we also consider superconductors with TRS. Naturally, TRS implies that
\bea
\hat{T}\hat\Delta\hat{T}^{-1}=\hat\Delta.
\eea
Using Eq.(\ref{eq:Delta}), we have
\bea\label{eq:CnTDelta1}
\hat{C}_n\hat{T}\hat\Delta\hat{T}^{-1}\hat{C}_n^{-1}=\hat{C}_n\hat\Delta\hat{C}_n^{-1}=e^{i2m\pi/n}\hat\Delta,
\eea
and
\bea\label{eq:CnTDelta2}
\hat{T}\hat{C}_n\hat\Delta\hat{C}_n^{-1}\hat{T}^{-1}=\hat{T}e^{i2m\pi/n}\hat\Delta\hat{T}^{-1}=e^{-i2m\pi/n}\hat{\Delta}.
\eea
However, since $[\hat{C}_n,\hat{T}]=0$, the $m$ which satisfies both Eq.(\ref{eq:CnTDelta1}) and Eq.(\ref{eq:CnTDelta2}) is $m=0,n/2$.

If $m=0$, $\hat{C}_{n,m}$ is the same as $\hat{C}_n$ and also commutes with time-reversal. When $m=n/2$, which is only possible if $n\in{even}$, the commutation relation between $\hat{C}_{n,n/2}$ and $\hat{T}$ is
\bea\label{eq:CnmandT}
\{C_{n,n/2},KT\}=0,
\eea
where $KT$ is the matrix representation of $T$ where $K$ is complex conjugation. Eq.(\ref{eq:CnmandT}) indicates that the ${C}_{n,n/2}$, $KT$ and $KP$ form a projective representation of the group generated by $T$, $C_n$ and $P$. We again defer the proof to Appendix \ref{app:PCT}.

\section{Classification of 1D Superconductors with Rotational Symmetries}
\label{sec:1d}
We now proceed to $1$D, and some quasi-$1$D systems, where the rotation along an axis parallel to the system is a symmetry represented by a matrix $C_{n,m}$ satisfying the commutation relation,
\bea
[C_{n,m},H(k)]=0.
\eea
It should be noted that in one dimension, there is no crystallographic constraint on $n$, and therefore $n\in{Z}^+$ (positive integers). As we classify the respective gapped superconductors with rotational symmetries in $1$D below, we separate the discussions into two distinct sections: one in which the total angular momentum of the Cooper pair is zero ($m=0$) and one where the angular momentum is non-zero ($m\neq0$).

\subsection{Pairing without Cooper Pair Angular Momentum ($m=0$)}

Most generic of BdG Hamiltonians, which belong to class D in the AZ classification table, have preserved particle-hole symmetry, which is represented by $KP$ in the Nambu basis, where $K$ is the complex conjugate and $P$ is a unitary matrix. Physically, we have $\hat{P}^2=1$ for an electron, which leads to
\bea\label{eq:PHS}
(KP)^2=PP^\ast=1,\\
\nonumber
P=P^T.
\eea

Since PHS changes electrons to holes and vice versa, it anti-commutes with the first quantized Hamiltonian and the momentum operator. This means that it sends one single particle state to another with opposite energy and momentum; symbolically we have
\bea
(KP)H(\bk)(KP)^{-1}=-H(-\bk),
\eea
or
\bea
PH(\bk)P^\dag=-H^T(-\bk).
\eea

In addition to this, PHS commutes with the rotation symmetry as
\bea\label{eq:PandCn}
[KP,C_n]=0\;\rightarrow\;PC_nP^\dag=C_n^\ast.
\eea
In the basis spanned by eigenstates of $C_n$, $(\phi_1,\phi_2,...)^T$, we have
\bea\label{eq:CnExpansion}
C_n=\sum_{\oplus{r}}rI_{d_r},
\eea
where $I_{d_r}$ is a $d_r$-by-$d_r$ identity matrix and $d_r$ is the degeneracy of the eigenvalue $r$. Using Eq.(\ref{eq:PandCn}) and Eq.(\ref{eq:CnExpansion}), we obtain the expression for $P$ in the basis $(\phi_1,\phi_2,...,\phi^\ast_1,\phi^\ast_2,...)^T$ as those spanned by the eigenstates.
\bea\label{eq:Pdecompose}
P=\sum_{\oplus{r\in{real}}}P_{r}\oplus\sum_{\oplus{\im[r]>0}}\left(\begin{matrix} 
      0 & Q_r \\
      Q_r^T & 0 \\
   \end{matrix}\right),
\eea
where $P_r$ and $Q_r$ are unitary matrices. This indicates that the PHS operator leaves unchanged the eigenspace of the rotation matrix, $C_n$, with a real eigenvalue, but maps the eigenspace with a complex eigenvalue to its complex conjugate.

Using Eq.(\ref{eq:PHS}) and Eq.(\ref{eq:Pdecompose}), we then have for $r\in{Real}$,
\bea
P_rH_r(\bk)P^\dag_r=-H^T_r(-\bk),
\eea
and for $r\notin{Real}$, we have
\bea\label{eq:complexP}
&&\left(\begin{matrix} 
      0 & Q_r \\
      Q_r^T & 0 \\
   \end{matrix}\right)\left(\begin{matrix} 
      H_r(\bk) & 0 \\
      0 & H_{r^\ast}(\bk) \\
   \end{matrix}\right)\left(\begin{matrix} 
      0 & Q^\ast_r \\
      Q_r^\dag & 0 \\
   \end{matrix}\right)\\
   \nonumber&=&-\left(\begin{matrix} 
      H^T_{r}(-\bk) & 0 \\
      0 & H^T_{r^\ast}(-\bk) \\
   \end{matrix}\right)
\eea

According to the AZ classification, $H_r(\bk)$ belongs to class D if $r\in{Real}$ and to class A (i.e., no symmetry because PHS relates $r$ to $r^\ast$) if $r\notin{Real}$, while $H_r(\bk)\oplus{}H_{r^\ast}(\bk)$ again belongs to class D. In 1D, class D has a $Z_2$ classification whereas class A is trivial. Therefore, each $H_{r\in{Real}}$ possesses its own $Z_2$-index. Therefore, the question that remains to be answered is can $H_r(\bk)\oplus{H}_{r^\ast}(\bk)$ be $Z_2$ nontrivial? We argue that it is impossible by examination of a simple contradiction. Should $H_r(\bk)\oplus{H}_{r^\ast}(\bk)$ be $Z_2$ nontrivial, then for an open chain there must be a single Majorana mode at each end\cite{Kitaev2001}. Due to $C_n$-symmetry, the Majorana mode must either have a rotation eigenvalue of $r$ or $r^\ast$, but either choice breaks the inherent PHS. Based on this discussion, we find that the topological classification of a $C_n$-invariant 1D superconductor without additional symmetries is given by a set of $Z_2$ numbers from each $H_r(\bk)$ with $r\in{Real}$. Using $r^n=-1$, it is obvious that for $n=even$ all eigenvalues are complex, thus, the classification is always trivial. Meanwhile for $n=odd$, $r=-1$ is the only real eigenvalue of $C_n$, and the only topological number is the $Z_2$ number of $H_{-1}(\bk)$. It must be noted that since the full Hamiltonian belongs to class D, which also has $Z_2$ classification, the $Z_2$ number of $H_{-1}(\bk)$ is the same as the $Z_2$ number of the full Hamiltonian.

In addition to PHS, which is shared by all superconductors, we consider the presence of TRS, corresponding to class DIII in the AZ classification. In the Nambu basis, TRS is represented by $KT$, where $T$ is a unitary matrix. For spinful electron, we have
\bea\label{eq:25}
(KT)^2=-1\rightarrow
T=-T^T.
\eea
The action of TRS reverses the momentum of an electron without changing its energy, or symbolically,
\bea\label{eq:TRS}
(KT)H(\bk)(KT)^{-1}=H(-\bk),\\\nonumber
TH(\bk)T^\dag=H^T(-\bk).
\eea
TRS also commutes with all spatial symmetries. Specifically,
\bea
[KT,C_n]=0\;\rightarrow\;TC_nT^\dag=C_n^\ast.
\eea
In the basis spanned by the eigenvectors of $C_n$, we have the following block-diagonalization of $T$
\bea\label{eq:Tdecompose}
T=\sum_{\oplus\im[r]=0}T_r\oplus\sum_{\oplus\im[r]>0}\left(\begin{matrix} 
      0 & R_r \\
      -R^T_r & 0 \\
   \end{matrix}\right),
\eea
where $T_r$ and $R_r$ are unitary matrices and $T_r$ is anti-symmetric from Eq.(\ref{eq:25}). Using Eq.(\ref{eq:TRS}) and Eq.(\ref{eq:Tdecompose}), we have for each $r\in{Real}$,
\bea\label{eq:realT}
T_rH_r(\bk)T_r^\dag=H^T_r(-\bk),
\eea
and for each $r\notin{Real}$
\bea\label{eq:complexT}
\left(\begin{matrix} 
      0 & R_r \\
      -R^T_r & 0 \\
   \end{matrix}\right)\left(\begin{matrix} 
      H_r(\bk) & 0 \\
      0 & H_{r^\ast}(\bk) \\
   \end{matrix}\right)\left(\begin{matrix} 
      0 & -R^\ast_r \\
      R^\dag_r & 0 \\
   \end{matrix}\right)\\\nonumber=\left(\begin{matrix} 
      H^T_r(-\bk) & 0 \\
      0 & H^T_{r^\ast}(-\bk) \\
   \end{matrix}\right).
\eea
From Eq.(\ref{eq:realT}), we understand that each sector with real $r$ has both TRS and PHS and hence belongs to class DIII, which in 1D has a $Z_2$ number. For $r\notin{Real}$, by utilizing a combination of Eq.(\ref{eq:complexP}) and Eq.(\ref{eq:complexT}), we obtain
\bea\label{eq:PT}
\left(\begin{matrix} 
      Q^\ast_rR_r^T & 0 \\
      0 & Q^\dag_rR_r \\
   \end{matrix}\right)\left(\begin{matrix} 
      H_r(\bk) & 0 \\
      0 & H_{r^\ast}(\bk) \\
   \end{matrix}\right)\left(\begin{matrix} 
      R_r^\dag{Q}_r^T & 0 \\
      0 & R_r^\dag{Q}_r \\
   \end{matrix}\right)\\\nonumber=-\left(\begin{matrix} 
      H_r(\bk) & 0 \\
      0 & H_{r^\ast}(\bk) \\
   \end{matrix}\right).
\eea
If we define $S_r=Q^\ast_rR_r^T$ and $S_{r^\ast}=Q^\dag_rR_r$ for each $r\notin{Real}$, Eq.(\ref{eq:PT}) leads to
\bea\label{eq:chiral}
\{S_r,H_r(\bk)\}=0,
\eea
which indicates that $H_{r\notin{Real}}(\bk)$ belongs to the chiral class AIII. Although a real sector $H_{r\in{Real}}$ also has chiral symmetry defined as $S_r=P_rT_r$, one cannot use this symmetry for classification, as the topological invariants protected by $S_r$ are constrained to certain numbers, zero in this case, by the individual PHS or TRS symmetry. Class AIII in 1D has a $Z$ number, so each sector with complex $r$ has a $Z$ number denoted by $z^{(r)}$. In fact, we can further argue that $z^{(r)}=-z^{(r^\ast)}$. The chiral symmetry is the composition of TRS and PHS, thus satisfying $S^2=P^2T^2=-1$. This indicates that $S_r$ has eigenvalues $\pm{i}$. Any class AIII Hamiltonian having $z^{(r)}>0$ ($z^{(r)}<0$) means that there are $|z^{(r)}|$ edge states at each end of an open system that are eigenstates of $S$ with eigenvalue $+i$ ($-i$). But under TRS, an edge state having $C_n$ eigenvalue $r$ and $S_r$ eigenvalue $+i$ maps to another edge state having $C_n$ eigenvalue $r^{\ast}$ and $S_r$ eigenvalue $-i$, implying that
\bea\label{eq:constraintZ}
z^{(r)}=-z^{(r^\ast)},
\eea
as a result of which the number of independent $Z$ numbers is determined by one-half the number of complex eigenvalues of $C_n$. Eq.(\ref{eq:constraintZ}) also implies that for $r\in{Real}$, TRS sets this topological number to zero, as $r=r^\ast$. For $n\in{}even$, the full classification is given by $n/2$ integers (as all $C_n$ eigenvalues appear in complex pairs), and if $n\in{odd}$, it is given by one $Z_2$ number and $(n-1)/2$ integers (for all eigenvalues except $-1$ appear in pairs). Again, the $Z_2$ number of the full Hamiltonian is given by the same as the $Z_2$ number of $H_{-1}(k)$ when $n\in{odd}$, and is trivial if $n\in{even}$.

Finally, let us consider spinless electrons, or equivalently, adding spin-SU(2) symmetry. We can follow all the steps above to find the classifications, keeping in mind the distinction that for spinless electrons, we have
\bea
T&=&T^T,\\
\nonumber
P&=&-P^T,\\
\nonumber
C^n_n&=&1.
\eea
Using nearly identical calculations to those presented in this section, we may derive the following additional results: (i) Without TRS, any $H_r(\br)$ with real $r$ and any $H_r(\bk)\oplus{H}_{r^\ast}(\bk)$ with complex $r$ belong to class C and have a trivial classification. (ii) Additionally, without TRS, any $H_r(\bk)$ with complex $r$ belongs to class A and also has a trivial classification. (iii) With TRS, any $H_r(\br)$ with real $r$ and any $H_r(\bk)\oplus{H}_{r^\ast}(\bk)$ with complex $r$ belong to class CI, which has a trivial classification in 1D (iv) In the presence of TRS, any $H_r(\bk)$ with complex $r$ belongs to class AIII, having a $Z$ classification, under the constraint as outlined in Eq.(\ref{eq:constraintZ}). Therefore in class CI, for $n\in{even}$, there are $(n-2)/2$ integers to specify the topological state and for $n\in{odd}$ there are $(n-1)/2$ integers, corresponding to the number of conjugate pairs of complex eigenvalues of $C_{n,m}$.

\subsection{Pairing with Cooper Pair Angular Momentum ($m\neq0$)}

When the angular momentum of the Cooper pair is considered, namely when $m\neq0$, $C_n$ must be replaced by $C_{n,m}$ as the rotation symmetry of the system. In the Nambu basis, it is represented by
\bea\label{eq:Cnm}
C_{n,m}=C_ne^{i\tau_z\frac{m\pi}{n}},
\eea
where $\tau_z$ is the Pauli matrix in the particle-hole indices due to $U_m$. We understand that PHS commutes with $C_{n,m}$ because (i) PHS commutes with $C_n$ and (ii) it anti-commutes with both $i$ and $\tau_z$, or
\bea\label{eq:PandCnm}
[KP,C_{n,m}]=0\;\rightarrow\;PC_{n,m}P^\dag=C^\ast_{n,m}.
\eea
Comparing Eq.(\ref{eq:PandCnm}) and Eq.(\ref{eq:PandCn}), we see that all preceding understanding obtained in the previous section where we ignored the angular momentum of the Cooper pair also applies to the system with both PHS and $C_{n,m}$ symmetry. Therefore, we simply apply the results and arrive at the following conclusions: (i) For $m,n\in{even}$, the classification of the system is trivial, as we have already stated. (ii) For $m\in{odd}$ and $n\in{even}$, $r$ may take the value $r=\pm1$, and $H_{\pm1}(\bk)$ belongs to class D and gives two $Z_2$ numbers, $z_2^{(\pm1)}$. (iii) For $m\in{even}$ and $n\in{odd}$, $r$ may take $-1$ but not $+1$, and $H_{-1}(\bk)$ has a $Z_2$ number $z_2^{(-1)}$. (iv) Finally, for $m,n\in{odd}$, $r$ may take the value of $+1$ but not $-1$, and $H_{1}(\bk)$ gives a $Z_2$ number $z_2^{(1)}$. The $Z_2$-index for the full Hamiltonian, neglecting rotation symmetry, is the same as the sum of the $Z_2$-indices corresponding to each sector with a real eigenvalue of $C_{n,m}$.

Now we consider adding TRS to the system. In Sec.\ref{sec:project} we have shown that the only nonzero $m$ that is compatible with TRS is $m=n/2$ when $n\in{even}$. In this case, $C_{n,n/2}$ and $KT$ anti-commute, i.e.,
\bea\label{eq:CnandT2}
TC_{n,n/2}T^\dag=-C^\ast_{n,n/2}.
\eea
Eq.(\ref{eq:CnandT2}) indicates that TRS maps a state with $C_{n,m}$ eigenvalue $r$ to a state with eigenvalue $-r^\ast$. We note that in this case time-reversal operator anti-commutes with $C_{n,m}$ from Eq.(\ref{eq:Cnm}) due to the fact that: (i) TRS commutes with $C_n$ yet anti-commutes with imaginary unit $i$ and (ii) it commutes with $\tau_z$ for it does not interchange particles and holes. Therefore, for the case when $m=n/2$ and we have $n\in{even}$, we separately discuss the following two constraints on the rotational symmetry: (i) $m\in{even}\Leftrightarrow{n}=4k$ and (ii) $m\in{odd}\Leftrightarrow{n}=4k-2$.  When $n=4k$ ($k$ being a non-negative integer) and the general eigenvalue of $C_{n,m}$ is $r_s=e^{i2\pi(s+1/2)/n}$ ($s$ being a non-negative integer), then under PHS the $r_s$-sector and the $r_{n-s-1}$-sector are mapped to each other while, at the same time, under TRS, the $r_s$-sector and the $r_{n/2-s-1}$-sector are mapped to each other. Therefore, $H_{r_s}\oplus{H}_{r_{n/2+s}}$ (note that $r_{n/2+s}=-r_{s}$ in this case) belongs to class AIII , thereby having a $Z$-index. (Since $P*T$ maps $s\rightarrow{n}-s-1\rightarrow{n/2}-(n-s-1)-1=n/2+s$, the direct sum is invariant under the composite symmetry.) However, this $Z$-number must vanish and this can be shown by contradiction. Suppose this $Z$ number is $z>0$, then on the edge there are $z$ states that are eigenvectors of $P*T$ having eigenvalue $+i$, such that $P*T$ in the Hilbert space spanned by the zero modes is $iI$, where $I$ is the identity matrix the dimension of which is the number of zero modes. Then we assume that the rotation symmetry $C_{n,m}$ be represented by some matrix $R$, and we have $[R,P*T]=0$ contradicting Eq.(\ref{eq:PandCnm},\ref{eq:CnandT2}). Considering the second case, $n=4k-2$, we know that the general eigenvalue of $C_{n,n/2}$ is $r_s=e^{2s\pi/n}$. Under PHS, the $r_s$-sector and the $r_{n-s}$ sectors are mapped to each other with the exceptions of $s=n/2$ and $s=n$, where $r=-1$ and $r=+1$, respectively. As in the previous case, under TRS, the $r_s$-sector and the $r_{n/2-s}$-sector are mapped to each other. Therefore, for $s\neq{n/2}$, the Hamiltonian $H_{r_s}\oplus{H}_{r_{n/2+s}}$ belongs to class AIII, while $H_{+1}\oplus{H}_{-1}$ belongs to class DIII. Based on above argument, the $Z$-index of the class AIII component must vanish, leaving to an overall classification to be $Z_2$. The $Z_2$-index for the full Hamiltonian, again without considering rotation symmetry, is the same as the $Z_2$-index for $H_{+1}\oplus{H}_{-1}$.

Finally, we consider adding SU(2) symmetry. With PHS and SU(2) the full Hamiltonian is in class C, having $(KP)^2=-1$ as the only symmetry. In this case, all sectors of the Hamiltonian belong either to class C or to class A, both being trivial. When we include TRS along with PHS and SU(2), we again need to separately consider the two cases above with regards to the rotational symmetry of the given system. We first consider the case for $n=4k$ in which a generic eigenvalue of $C_{n,n/2}$ is $r_s=e^{2s\pi/n}$. Under PHS, the $r_s$-sector and the $r_{n-s}$ sectors are mapped to each other with the exceptions of $s=n/2$ and $s=n$. Further, under the application of TRS, the $r_s$-sector and the $r_{n/2-s}$-sector are mapped to each other, with the exceptions of $s=n/4$ and $s=3n/4$, where $r_{s}$-sector is mapped to itself. Therefore, for $s\neq{n/4},{n/2},3n/4,n$, the Hamiltonian $H_{(r_s)}\oplus{H}_{(r_{n/2+s})}$ belongs to class AIII and is characterized by a vanishing $Z$-index. Similarly, the sectors corresponding to $H_{(+1)}\oplus{H}_{(-1)}$ and $H_{(+i)}\oplus{H}_{(-i)}$ belong to class CI, and possess only a trivial classification resulting in an overall classification for the system that is trivial. For the case corresponding to rotational symmetries satisfying $n=4k-2$, we have eigenvalue of $r_s=e^{i2\pi(s+1/2)/n}$. As before, under PHS, the $r_s$-sector and the $r_{n-s-1}$-sector are mapped to each other, and the application of TRS maps the $r_s$-sector and the $r_{n/2-s-1}$-sector to one another, with the exception of $s=(n/2-1)/2$. Therefore, for $s\neq(n/2-1)/2,(n/2+1)/2$, the Hamiltonian $H_{r_s}\oplus{H}_{r_{n/2+s}}$ belongs to class AIII, having a $Z$-classification but with vanishing $Z$-index, while $H_{(+i)}\oplus{H}_{(-i)}$ belongs to class CI, having trivial classification. Accordingly, the overall classification is again trivial.

In Table \ref{tab:1D}, We summarize the classification of all gapped 1D superconductors within four BdG classes (C, D, CI and DIII) of the AZ classification enhanced by $C_{n,m}$-symmetry.

\begin{table}[htp]
\begin{center}
\begin{tabular}{|c|c|c|c|}
\hline
 & $n\in{odd}$ & $n=4k$ & $n=4k-2$\\
\hline
C & 0 & 0 & 0\\
\hline
D, $m\in{even}$ & 0 & $Z_2$ & $Z_2$\\
\hline
D, $m\in{odd}$ & $Z_2^2$ & $Z_2$ & $Z_2$\\
\hline
CI, $m=0$ & $Z^{(n-1)/2}$ & $Z^{(n-2)/2}$ & $Z^{(n-2)/2}$\\
\hline
CI, $m=n/2$ & 0 & 0 & 0\\
\hline
DIII, $m=0$ & $Z_2\times{Z}^{(n-1)/2}$ & $Z^{n/2}$ & $Z^{n/2}$\\
\hline
DIII, $(m=n/2)$ & $0$ & 0 & $Z_2$\\
\hline
\end{tabular}
\end{center}
\caption{Complete classification of 1D gapped superconductors with rotation symmetry, $C_{n,m}$. Within the table, `0' indicates that for the given system the classification is trivial.\label{tab:1D}}
\end{table}%

\section{2D Superconductors with Rotation Symmetries}
\label{sec:2d}

With our discussion of gapped $1$D superconductors with rotation symmetries complete, we focus our attention in this section on the study of $2$D gapped superconductors without TRS (class D and class C)in the presence of $C_{n,m}$ invariance, where the rotation axis is assumed to be the axis perpendicular to the system. Lattice periodicity is compatible with rotation symmetry only when $n=2,3,4,6$\cite{Scherrer1946} . In BZ, there exist high-symmetry points that are invariant under $C_{\tilde{n}}$, where $\tilde{n}$ is a factor of $n$, denoted by $\mathbf{K}_{\tilde{n}}$. For example, in a $C_4$-invariant system, $X=(\pi,0)$ and $Y=(0,\pi)$ are points that are $C_2$ invariant; and in a $C_6$-invariant system, $K$ and $K'$ are $C_3$-invariant. At $\mathbf{K}_{\tilde{n}}$, each energy eigenstate is also an eigenstate of $C_{\tilde{n},{m}}$ (it being understood that $m$ is a mod $\tilde{n}$ number). For each eigenvalue $r$ of $C_{\tilde{n},{m}}$, we count at $\mathbf{K}_{\tilde{n}}$ the number of occupied energy eigenstates that are also eigenstates of $C_{\tilde{n}{m}}$ with eigenvalue $r$ and denote it by $N_r(\bK_{\tilde{n}})$. We show that these numbers are related to the Chern numbers in superconductors. Physically, the Chern number of a superconductor is determined by both the band structure of the normal state and the symmetry of the pairing amplitude on the Fermi surface. The former contribution is related to the $C_n$ eigenvalues of the occupied bands at high symmetry points in the normal states\cite{Fang2012,Chiu2014tcs,Hughes2014tcs,Teo2013tcs}, while the latter contribution is related to $m$, namely, the angular momentum of the Cooper pair modulo $n$. In this section, we focus on how these numbers relate to the Chern number in gapped $2$D superconductors.


\subsection{Continuum Limit ($n=\infty$)}

To begin our analysis, let us first consider the continuum limit with full SO(2) symmetry. In this limit, the angular momentum of the Cooper pair, $m$, can take any integer. We choose to work in an orbital basis in which the generator of the rotation operator $\hat{J}$ is diagonalized. Therefore, in the Nambu basis, we have
\bea
\tilde{J}=\tau_z\otimes{}diag\{j_1,j_2,...,j_{N_{orb}}\},
\eea
where $\tau_z$ is the Pauli matrix acting on the particle-hole index and $j_\alpha$ is the angular momentum of the $\alpha$-th electronic orbital. The second quantized form of $\hat{J}$ is given by
\bea
\hat{J}=\sum_{i=1,...,N_{orb}}j_i\phi_i^\dag\phi_i,
\eea where $\phi_i$ is the annihilation operator of angular momentum $j_i$. Under rotation through $\theta$ via the application of the rotation operator, we have
\bea\label{eq:45}
e^{i\hat{J}\theta}\hat\Delta{e}^{-i\hat{J}\theta}=e^{im\theta}\hat\Delta,
\eea
or its infinitesimal version
\bea\label{eq:inf45}
[\hat{\Delta},\hat{J}]=-m\hat{\Delta}.
\eea
Furthermore, we also know
\bea\label{eq:taudeltacom}
[\hat{\Delta},\hat{Q}]=-2\hat{\Delta},
\eea
where $\hat{Q}\sum{c}^\dag_\alpha{c}_\alpha$ is the total charge. Using Eq. (\ref{eq:taudeltacom}) in conjunction with Eq. (\ref{eq:inf45}), we can prove that
\bea\label{eq:42}
[\hat{J}_m,\hat\Delta]=0,
\eea
where $\hat{J}_m\equiv\hat{J}-\frac{m}{2}\prod_{\alpha}(1-2c^\dag_\alpha{c}_\alpha)$. In the Nambu basis, $\hat{J}_m$ is represented by
\bea
\tilde{J}_m=\tilde{J}-\frac{m}{2}\tau_z\otimes{I}_{N_{orb}}.
\eea
Since $\hat{J}$ commutes with the normal part, $\hat{H}_0$ of the Hamiltonian, using Eq.(\ref{eq:42}), we know that $\hat{J}_m$ commutes with the full Hamiltonian
\bea
[\hat{J}_m,\hat{H}]=0,
\eea
or, in the presence of translational symmetry
\bea
\begin{split}
H(k_+e^{i\theta},k_-e^{-i\theta})=\exp(i(\tilde{J}-\frac{m}{2}\tau_z)\theta) \times \\ H(k_+,k_-)\exp(-i(\tilde{J}-\frac{m}{2}\tau_z)\theta),
\end{split}
\eea
where $k_\pm=k_x\pm{i}k_y$.
In the 2D continuum k-space, $\bk=0$ and $\bk=\infty$ are the only two points that are invariant under rotation. At these points we have
\bea
[\tilde{J}_m,H(0)]=[\tilde{J}_m,H(\infty)]=0,
\eea
where we have implicitly assumed that $H(\infty)$ is well defined. Each state of $H(0)$ or $H(\infty)$ is also an eigenstate of $\tilde{J}_m$ of eigenvalue $j^i_m$ ($i$ denoting the occupied bands in the BdG Hamiltonian). In Appendix \ref{app:cproof}, we prove a general relation between the Chern number and all $j_m$'s at $\bk=0$ and $\bk=\infty$:
\bea\label{eq:Chern}
C=\sum_{i=1}^{N_{orb}}[j_m^i(0)-j_m^i(\infty)].
\eea
If one considers the gapped BdG Hamiltonian to be the same as that of an insulator with accidental particle-hole symmetry, Eq.(\ref{eq:Chern}) simply means that its Chern number equals the total angular momentum (along $z$-axis), where $\hat{J}_m$ is the angular momentum operator, of all occupied states. To heuristically observe this, we notice that for any occupied state, $|\psi(\bk)\rangle$ at a generic $\bk\neq0,\infty$, the state $e^{i\hat{J}_m\theta}|\psi(\bk)\rangle$ must also be an occupied state with momentum $R(\theta)\bk$. One can always construct $|\psi_n\rangle=\int_0^{2\pi}d\theta{}e^{in\theta}e^{i\hat{J}_m\theta}|\psi(\bk)\rangle$ for any integer $n$ and, therefore, all contribution to the total angular momentum from generic $\bk$'s cancel each other, leaving the only contribution from $\bk=0,\infty$. We then recall that any rotation about $\bk=0$ is equivalent to an inverse rotation about $\bk=\infty$, so the total angular momentum is the difference, not the sum, of $j^i_m$'s at $0$ and $\infty$.

Now examine the weak pairing limit, where we may separate the contribution due to the normal state band structure from that of the pairing on the Fermi surface. In the weak coupling limit, at each $\bk$, the occupied bands in the BdG Hamiltonian consist of two distinct parts: the occupied bands of the non-superconducting Hamiltonian, and the particle-hole partner of all the unoccupied bands. Keeping in mind that a hole state has opposite charge and angular momentum compared with an electron state, we have
\bea\label{eq:weak}
\sum_{i=1}^{N_{orb}}j_m^i(\bK)&=&j^1(\bK)+...+j^{N_{occ}(\bK)}(\bK)\\
\nonumber
&-&(j^{N_{occ}(\bK)+1}(\bK)+...+j^{N_{orb}}(\bK))\\
\nonumber
&-&\frac{m}{2}[N_{occ}(\bK)-N_{unocc}(\bK)],
\eea
where $N_{occ}$ and $N_{unocc}$ are the number of occupied and unoccupied bands, respectively.
Substituting Eq.(\ref{eq:weak}) to Eq.(\ref{eq:Chern}), we obtain a simple formula
\bea\label{eq:simpleChern}
C=2[J(0)-J(\infty)]-m[N_{occ}(0)-N_{occ}(\infty)].
\eea
The physical meaning of Eq.(\ref{eq:simpleChern}) is clear as the first term is simply two times the total angular momentum of the normal state, where the factor of two is because of the Fermion doubling in the Nambu basis. The second term is the total angular momentum of the pairing on all Fermi surfaces. To see this, we need to notice two separate facts: (i) Eq.(\ref{eq:45}) indicates that $m$ is the total angular momentum of a Cooper pair, and (ii) $N_{occ}(0)-N_{occ}(\infty)$ is the difference in the occupation numbers at $\bk=0$ and $\bk=\infty$. Suppose $N_{occ}(0)>N_{occ}$, and by traversing any path from $0$ to $\infty$ one crosses $N_e$ electron-like Fermi surfaces and $N_h$ hole-like Fermi surfaces, then we have $N_{e}-N_h=N_{occ}(0)-N_{occ}(\infty)$.

\subsection{Finite Rotational Symmetry $C_n$ ($n=2,3,4,6$)}

When considering the more realistic case of a 2D lattice, the continuous rotation symmetry $C_{\infty}$ breaks down to $C_{n=2,3,4,6}$. In our approach here, we closely follow our previous work\cite{Fang2012}, in order to obtain the Chern number up to a multiple of $n$ in terms of the eigenvalues of $C_{\tilde{n},m}$ at k-points invariant under $C_{\tilde{n}}$ where $\tilde{n}$ divides $n$. For $C_{n=2,3,4,6}$ we find:
\bea\label{eq:general}
e^{i2\pi{C}/2}=\zeta_m(\Gamma)\zeta_m(M_1)\zeta_m(M_2)\zeta_m(M_3),\\
\nonumber
e^{i2\pi{C}/3}=(-1)^{N_{orb}(m+1)}\theta_m(\Gamma)\theta_m(K)\theta_m(K'),\\
\nonumber
e^{i2\pi{C}/4}=(-1)^{N_{orb}(m+1)}\xi_m(\Gamma)\xi_m(M)\zeta_m(X),\\
\nonumber
e^{i2\pi{C}/6}=(-1)^{N_{orb}(m+1)}\eta_m(\Gamma)\theta_m(K)\zeta_m(M),
\eea
where $\zeta_m,\theta_m,\xi_m,\eta_m$ are the product of all eigenvalues of $C_{2,m}$, $C_{3,m}$, $C_{4,m}$ and $C_{6,m}$ at corresponding high-symmetry points on the lower half BdG bands, respectively. In the weak coupling limit, they again reduce to expressions that only involve the eigenvalues of $C_{\tilde{n}}$, the occupation number at each high-symmetry point and the angular momentum of the Cooper pair (mod $n$). To be specific,
\begin{widetext}
\bea\label{eq:CnChern}
e^{i2\pi{C}/2}&=&\exp[i(m+1)\pi(N_{occ}(\Gamma)+N_{occ}(M_1)+N_{occ}(M_2)+N_{occ}(M_3)],\\
\nonumber
e^{i2\pi{C}/3}&=&\frac{\theta^2(\Gamma)}{\theta(K)\theta(K')}\exp[i\frac{2m\pi}{3}(2N_{occ}(\Gamma)-N_{occ}(K)-N_{occ}(K'))],\\
\nonumber
e^{i2\pi{C}/4}&=&\frac{\xi^2(\Gamma)}{\xi^2(M)}\exp(-i\frac{2m\pi}{4})(N_{occ}(\Gamma)+N_{occ}(M)-2N_{occ}(X)),\\
\nonumber
e^{i2\pi{C}/6}&=&\frac{\eta^2(\Gamma)\zeta(M)}{\theta(K)}\exp[-i\frac{2m\pi}{6}(N_{occ}(\Gamma)+2N_{occ}(K)-3N_{occ}(M))].
\eea
\end{widetext}
and the definition of high-symmetry points is given in Fig.\ref{fig:SBZ}. Let us derive the $n=2$ case here in detail. In the weak coupling limit, each occupied state of $C_2$ eigenvalue $\zeta$ at $\Gamma$ is an eigenstate of $C_{2,m}$ with eigenvalue $\zeta{e}^{i\frac{m\pi}{2}}$, and each unoccupied state with eigenvalue $\zeta_j$ at $\Gamma$ is an eigenstate of $C_{2,m}$ with eigenvalue $\zeta_j^\ast{e}^{-i\frac{m\pi}{2}}$ after PHS transform. Therefore, the total product of eigenvalues of $C_{2,m}$ at $\Gamma$ is
\bea
\zeta_m(\Gamma)&=&(\prod_{i\in{occ}}\zeta_i{e}^{i\frac{m\pi}{2}})(\prod_{j\in{unocc}}\zeta^\ast_j{e}^{i\frac{-m\pi}{2}})\\
\nonumber
&=&(\prod_{i\in{occ}}\zeta_i)^2(\prod_{n\in{orb}}\zeta^\ast_n)e^{im\pi{N}_{occ}(\Gamma)}e^{-i\frac{m\pi}{2}N_{orb}}.
\eea

If $m=0$, then we have (since $C_2^2=-1$)
\bea
(\prod_{i\in{occ}}\zeta_i)^2=(-1)^{N_{occ}(\Gamma)},
\eea
so\begin{widetext}
\bea
(-1)^C&=&(-1)^{N_{occ}(\Gamma)+N_{occ}(M_1)+N_{occ}(M_2)+N_{occ}(M_3)}(\prod_{n\in{orb}}\zeta^\ast_n)^4\\
\nonumber
&=&(-1)^{N_{occ}(\Gamma)+N_{occ}(M_1)+N_{occ}(M_2)+N_{occ}(M_3)}
\eea
\end{widetext}
If $m=1$, then
\bea
(\prod_{i\in{occ}}\zeta_i)^2=1,
\eea
so\begin{widetext}
\bea
(-1)^C&=&(-1)^{2N_{occ}(\Gamma)+2N_{occ}(M_1)+2N_{occ}(M_2)+2N_{occ}(M_3)}(\prod_{n\in{orb}}\zeta^\ast_n)^4e^{-i2{\pi}N_{orb}}\\
\nonumber
&=&1.
\eea
\end{widetext}

In Eq.(\ref{eq:CnChern}), the contribution to the Chern number again decomposes into two parts as promised, but the physical meaning is not as transparent as in Eq.(\ref{eq:simpleChern}), because here the angular momentum is well-defined only up to a multiple of $n$, and states at high-symmetry points other than $\Gamma$ contribute to the total angular momentum in different ways. We hope our eigenvalue formulas for projector Chern numbers can be useful in the search of topological chiral superconductors.

\section{3D Superconductors with Rotational Symmetries}
\label{sec:3d}

\subsection{Bulk Invariants}

\begin{figure*}
\includegraphics[width=16cm]{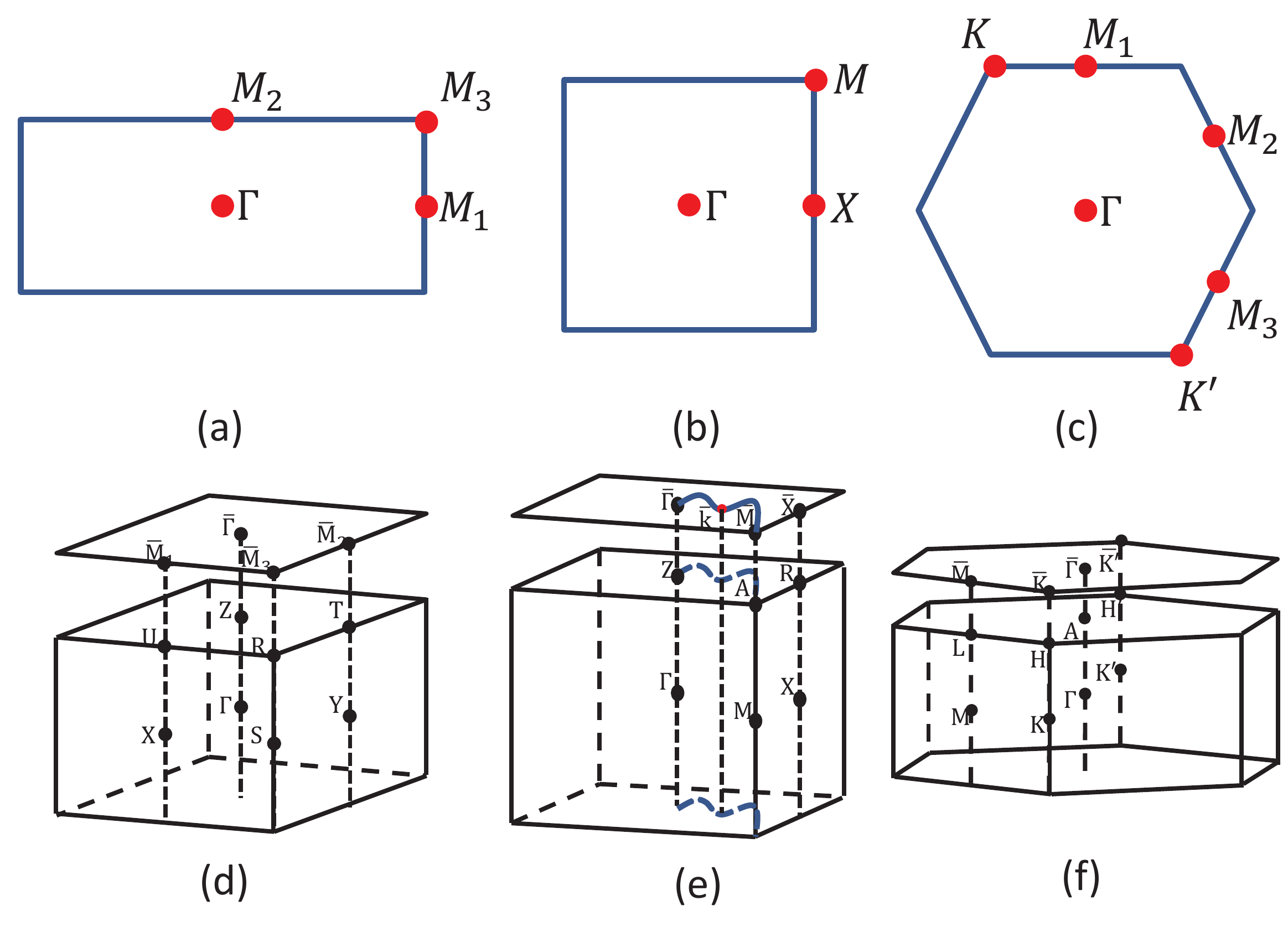}
\caption{(a,b,c) show the Brillouin zones of 2D systems with rotation symmetries, where (a) is for $n=2$, (b) for $n=4$ and (c) for $n=3,6$. (d,e,f) show the Brillouin zones of 3D systems with rotation symmetries and the surface Brillouin zones for any termination perpendicular to the rotation axis, where (d) is for $n=2$, (e) for $n=4$ and (f) for $n=3,6$.\label{fig:SBZ}}
\end{figure*}

We move on to discuss the partial topological classification of 3D gapped superconductors with rotation symmetries. In a 3D lattice, crystallographic constraint dictates that $n=2,3,4,6$ and within a $C_n$ invariant lattice system, there are discrete lines in the 3D BZ that are invariant under $C_{\tilde{n}>1}$ where $\tilde{n}$ is a factor of $n$. Therefore, in order to classify $3$D gapped superconducting systems, we can apply the classification of 1D superconductors with $C_{\tilde{n}}$ invariance to these sub-manifolds, and the set of quantum numbers of all $C_{\tilde{n}}$-invariant lines gives the enhanced classification of the 3D system. We notice that for $\tilde{n}=2$, $4$, and $6$ all three local symmetries we have discussed in the text, namely time-reversal, particle-hole and spin rotation, of the 3D system are also preserved on the $C_{\tilde{n}}$ invariant lines. The same applies to the line that includes $\Gamma$ when $n=3$. In Fig.\ref{fig:SBZ}, we schematically represent these special lines by their projections onto the surface BZ, where the surface is perpendicular to the rotation axis. For the outset, we show that special treatment is needed for the following lines: (i) a generic vertical (parallel to the rotation axis) line in BZ in a system with $C_{2,4,6}$-symmetry and (ii) a $C_3$-invariant line that does not include $\Gamma$. We must treat these lines specially because, while TRS and PHS are not symmetries, compositions such as $C_2*P$ and $C_2*T$, $P*T$ might be symmetries. Due to the lack of PHS on these generic lines, the classification derived in Sec.\ref{sec:1d} does not apply for these lines.
In a system with $C_{2,4,6}$-symmetry, a generic vertical line does not have TRS or PHS, but it may have the following symmetries: $KP'\equiv{KP}C^{n/2}_{n,m}$ and $KT'\equiv{KT}C^{n/2}_{n,m}$, where we should note that TRS is only possible for $m=0$ and $m=n/2$. To see this, we note that each generic vertical line is labeled by its momentum perpendicular to the rotation axis, $\bar\bk$, and under either $T$, $P$ or $C_{n,m}^{n/2}$, this momentum is mapped to $-\bar\bk$, but under $P'$ or $T'$ the momentum is mapped back to $\bar\bk$, while sending the momentum along the line to its opposite value. Using the commutation relations similar in form to those used in Eq. (\ref{eq:PandCnm}) and Eq. (\ref{eq:CnmandT}) , we have
\bea\label{eq:newsym}
({KP'})^2&=&(-1)^{m+F}(KP)^2=(-1)^{m+1},\\
\nonumber
({KT'})^2&=&(-1)^{mn/2+m+F}(KT)^2=(-1)^F(KT)^2=1.
\eea
Eq.(\ref{eq:newsym}) makes the class of generic lines different from the class of the whole system, depending on the parity of $m$. Let us examine each of the subsequent possibilities of lines embedded in the 3D BZ in turn: Case (i): class C and $m$ even, then Eq.(\ref{eq:newsym}) states that $({KP'})^2=(KP)^2=-1$, thus $H(\bar\bk)$ belongs to class C and has trivial classification. Case (ii): class C and $m$ odd, then Eq.(\ref{eq:newsym}) states that $({KP'})^2=-(KP)^2=1$, so $H(\bar\bk)$ belongs to class D, which gives it a $Z_2$ classification. Case (iii): class D and $m$ even, then Eq.(\ref{eq:newsym}) states that $({KP'})^2=-(KP)^2=-1$ indicating that $H(\bar\bk)$ belongs to class C and has a trivial classification. Case (iv): class D and $m$ odd, Eq.(\ref{eq:newsym}) states that $({KP'})^2=(KP)^2=1$, so $H(\bar\bk)$ belongs to class D, having $Z_2$ classification. Case (v): class CI and $m$ even and Eq.(\ref{eq:newsym}) states that $({KP'})^2=(KP)^2=-1$ and $({KT'})^2=(KT)^2=1$, so $H(\bar\bk)$ belongs to class CI, having trivial classification. Case (vi): class CI and $m$ odd thus Eq.(\ref{eq:newsym}) states that $({KP'})^2=-(KP)^2=1$ and $({KT'})^2=(KT)^2=1$, indicating that $H(\bar\bk)$ belongs to class BDI, which has a $Z$ classification. Case (vii): class DIII and $m$ even so Eq.(\ref{eq:newsym}) states that $({KP'})^2=-(KP)^2=-1$ and $({KT'})^2=-(KT)^2=1$ indicating that $H(\bar\bk)$ belongs to class CI, having trivial classification. Case (viii): class DIII and $m$ odd, Eq.(\ref{eq:newsym}) states that $({KP'})^2=(KP)^2=1$ and $({KT'})^2=-(KT)^2=1$, so $H(\bar\bk)$ belongs to class BDI and has a $Z$ classification.

We now consider a $C_3$-invariant line that does not include $\Gamma$. If the system under consideration has $C_3$-invariance but not $C_6$-invariance, then the line does not possess PHS. The Hamiltonian on the line hence belongs to class A and accordingly possesses a trivial classification. However, if the system also contains TRS , the line possesses the combined symmetry $KP*KT$. Since both $KT$ and $KP$ commute with $C_3$, each sector of the occupied state has the chiral symmetry or,
\bea\label{eq:34}
\{H_{r},S_r\}=0,
\eea
where the eigenvalues are $r\in(-1)^F\{1,\omega\equiv{e}^{i2\pi/3},\bar\omega\equiv{e}^{-i\pi/3}\}$. $S_r$ here is the matrix representation of the chiral symmetry $S=P*T$ projected to the $r$-sector. Eq.(\ref{eq:34}) indicates that each sector is in class AIII, and since class AIII has $Z$-classification, one $C_3$-invariant line has $Z^3$-classification. Further, we notice that one $C_3$-invariant line that does not include $\Gamma$ is mapped to another $C_3$-invariant line under either the application of either TRS or PHS. TRS, or equivalently PHS, flips the sign of the invariant in each sector as it sends a state with $S=+i$ to a state with $S=-i$, and maps the sector with $r=\pm\omega$ to the one with $r=\pm\bar\omega$. Therefore, if one line has topological number $(z^{(\pm\omega)},z^{(\pm\bar\omega)},z^{(\pm1)})=(z_1,z_2,z_3)$, the topological number of the other line is fixed to be $(z^{(\pm\omega)},z^{(\pm\bar\omega)},z^{(\pm1)})=(-z_2,-z_1,-z_3)$. Furthermore, if a superconductor has $C_{6,m}$-symmetry, the $C_3$-invariant line also possesses $KP'$ and if the system has time-reversal, it posseses $KT'$. The class of the Hamiltonian on this type of line depends on the class of the system and the parity of $m$, determined via Eq.(\ref{eq:newsym}). As long as the class of the line is one of the four classes that we have discussed, C, D, CI or DIII, one can simply use Table \ref{tab:1D} to obtain the $C_3$ enhanced classification. However, a separate analysis is needed for case (vi) (class CI and $m\in{odd}$), and case (viii) (class DIII and $m\in{odd}$), where the line belongs to class BDI. With $C_3$-symmetry, the $r=\pm1$ sector is invariant under both $KP'$ and $KT'$, and therefore belongs to class BDI, which has a $Z$-index. The sectors with $r=\pm\omega$ and $r=\pm\bar\omega$ are invariant under $S'=KP'*KT'$ and are mapped to each other under $KP'$ or $KT'$. Therefore each of them is in class AIII, which has a $Z$-index, and the two indices are related to each other by either $KP'$ or $KT'$, therefore, the overall classification becomes $Z^2$.

We move on to discuss the classifications of other high-symmetry lines. The high-symmetry lines are separated in $\bk$-space, but their quantum numbers are not independent of each other in a fully gapped 3D superconductor. This is because the special lines can be adiabatically connected to each other by an interpolation consisting of a series of generic lines which are \emph{not} invariant under rotations or time-reversal. In Fig.\ref{fig:SBZ}(e), for example, $\Gamma{Z}$ may continuously move to $MA$ by shifting the two end points along the dotted paths in the BZ, through a series of generic lines such as the dotted line that projects onto $\bar\bk$. The fully gapped bulk provides relations between the invariants obtained on high-symmetry lines and a generic line and the relationship may be summarized in two simple yet general rules. The first rule states that if a generic line has trivial classification, the topological invariants at high-symmetry lines are not related to (constrained by) each other. It is supported by the following argument without a rigorous proof. Consider two high-symmetry generic lines, $L_{1,2}$ that have nontrivial topological invariants thus exhibiting $n_{1,2}$ Majorana bound states at high-symmetry points in the SBZ, $\bar{L}_{1,2}$, for a $C_n$-invariant termination. As $L_1$ moves to another high-symmetry line $L_2$, since the generic lines are trivial, the $n_1$ Majorana modes at $\bar{L}_1$ move away from zero energy at a generic $\bar\bk$ into the bulk, and as the path reaches $L_2$, $n_2$ modes emerge from the bulk to meet at $\bar{L}_2$ with no relation to the $L_1$ line. On the other hand, when the generic (vertical) lines have nontrivial classification, the topological invariant of a generic line gives constraints on those of high-symmetry lines (see below for an example). The second rule states that if any high-symmetry line has trivial classification, the topological invariant of a generic line (given that the presence of a nontrivial classification such that an invariant can be defined) must be zero (trivial). This can be proved by contradiction: if a generic line has Majorana modes at $\bar\bk$ at zero energy, these modes will remain at zero energy as the line moves in the BZ to any high-symmetry line, so the high-symmetry line would have nontrivial classification, against our assumption. For cases where both the generic and high-symmetry lines have a non-trivial classification, that is, when neither of the rules applies, special analysis is needed. In the following we will derive these relations for two cases in detail, and the complete result, obtained via identical methodology, is given in Table \ref{tab:3D}.

\begin{table*}[htp]

\begin{tabular}{|c|c|c|c|c|c|c|c|}
\hline
Class & $(n,m)$ & $\bar{k}$ & $\bar{M}_{1,2,3},\bar{X}$ & $\bar{K},\bar{K}'$ & $\bar{M}$ & $\bar{\Gamma}$ & Constraints by a bulk gap\\
\hline
C & $(2,0)$ & $0$ & $0$ & N/A & N/A & $0$ & None\\
\hline
C & $(2,1)$ & $Z_2$ & $0$ & N/A & N/A & $0$ & $z_2(\bar{k})=0$\\
\hline
C & $(3,0),(3,1),(3,2)$ & $0$ & N/A & $0$ & N/A & $0$ & None\\
\hline
C & $(4,0),(4,2)$ & $0$ & $0$ & N/A & $0$ & $0$ & None\\
\hline
C & $(4,1),(4,3)$ & $Z_2$ & $0$ & N/A & $0$ & $0$ & $z_2(\bar{k})=0$\\
\hline
C & $(6,0),(6,2),(6,4)$ & $0$ & $0$ & $0$ & N/A & $0$ & None\\
\hline
C & $(6,1),(6,3),(6,5)$ & $Z_2$ & $0$ & $Z_2$ & N/A & $0$ & $z_2(\bar{k})=z_2(\bar{K})=0$\\
\hline
D & $(2,0)$ & $0$ & $0$ & N/A & N/A & $0$ & None\\
\hline
D & $(2,1)$ & $Z_2$ & $Z_2\times{Z}_2$ & N/A & N/A & $Z_2\times{Z}_2$ & $z^{(+1)}_2(\bar{M}_{1,2,3},\bar\Gamma)+z_2^{(-1)}(\bar{M}_{1,2,3},\bar\Gamma)=z_2(\bar{k})$\\
\hline
D & $(3,0),(3,1),(3,2)$ & $0$ & N/A & $0$ & N/A & $Z_2$ & None\\
\hline
D & $(4,0),(4,2)$ & $0$ & $0$ & N/A & $0$ & $0$ & None\\
\hline
D & $(4,1),(4,3)$ & $Z_2$ & $Z_2\times{Z}_2$ & N/A & $Z_2\times{Z}_2$ & $Z_2\times{Z}_2$ & $z^{(+1)}_2(\bar{\Gamma},\bar{M},\bar{X})+z^{(-1)}_2(\bar{\Gamma},\bar{M},\bar{X})=z_2(\bar{k})$\\
\hline
D & $(6,0),(6,2),(6,4)$ & $0$ & $0$ & $0$ & N/A & $0$ & None\\
\hline
D & $(6,1),(6,3),(6,5)$ & $Z_2$ & $Z_2\times{Z}_2$ & $Z_2$ & N/A & $Z_2\times{Z}_2$ & $z^{(+1)}_2(\bar{\Gamma},\bar{M})+z^{(-1)}_2(\bar{\Gamma},\bar{M})=z_2(\bar{k})=z_2(\bar{K})$\\
\hline
CI & $(2,0)$ & $0$ & $0$ & N/A & N/A & $0$ & None\\
\hline
CI & $(2,1)$ & $Z$ & $0$ & N/A & N/A & $0$ & $z(\bar{k})=0$\\
\hline
CI & $(3,0)$ & $Z$ & N/A & $Z^3$ & N/A & $Z$ & $z(\bar{k})=z^{(+1)}(\bar{K})+z^{(\omega)}(\bar{K})+z^{(\bar\omega)}(\bar{K})=0$\\
\hline
CI & $(4,0)$ & $0$ & $0$ & N/A & $Z$ & $Z$ & None\\
\hline
CI & $(4,2)$ & $0$ & $0$ & N/A & $0$ & $0$ & None\\
\hline
CI & $(6,0)$ & $0$ & $0$ & $Z$ & N/A & $Z^2$ & None\\
\hline
CI & $(6,3)$ & $Z$ & $0$ & $Z^2$ & N/A & $0$ & $z(\bar k)=z^{(-1)}(\bar{K})+2z^{(-\omega)}(\bar{K})=0$\\
\hline
DIII & $(2,0)$ & $0$ & $Z$ & N/A & N/A & $Z$ & None\\
\hline
DIII & $(2,1)$ & $Z$ & $Z_2$ & N/A & N/A & $Z_2$ & $z(\bar{k})=0$\\
\hline
DIII & $(3,0)$ & $Z$ & N/A & $Z^3$ & N/A & $Z_2\times{Z}$ & $z(\bar{k})=z^{(-1)}(\bar{K})+z^{(-\omega)}(\bar{K})+z^{(-\bar\omega)}(\bar{K})=0$\\
\hline
DIII & $(4,0)$ & $0$ & $Z$ & N/A & $Z^2$ & $Z^2$ & None\\
\hline
DIII & $(4,2)$ & $0$ & $Z$ & N/A & $0$ & $Z_2$ & None\\
\hline
DIII & $(6,0)$ & $0$ & $Z$ & $Z$ & N/A & $Z^3$ & None\\
\hline
DIII & $(6,3)$ & $Z$ & $Z_2$ & $Z^2$ & N/A & $Z_2$ & $z(\bar k)=z^{(1)}(\bar{K})+2z^{(\omega)}(\bar{K})=0$\\
\hline
\end{tabular}
\label{tab:special}
\caption{The complete classification of 1D vertical lines in the BZ of a 3D superconductor with $C_{n,m}$-invariance. The column $\bar\bk$ contains the classification for a generic line in the bulk. All other columns except the last contain the classification of high-symmetry lines that project to high-symmetry points in the surface BZ. In the last column, we list the constraints between the invariants on these lines placed by a full superconducting gap in the 3D BZ. Within the table, '0' indicates a trivial entry while 'N/A' indicates that the particular constraints considered in a particular tabular entry are not applicable.}\label{tab:3D}
\end{table*}%

For our first example, we consider a class D Hamiltonian with $C_{2,1}$-invariance. At the high-symmetry lines in BZ, $\Gamma{Z}$ projecting to $\bar\Gamma$, $XU$ projecting to $\bar{M}_1$, $YT$ projecting to $\bar{M}_2$ and $SR$ projecting to $\bar{M}_3$ in the surface BZ [see Fig.\ref{fig:SBZ}(d)], the Hamiltonian has $Z_2\times{Z}_2$ classification. The two $Z_2$-indices correspond to the two sectors with $C_{2,1}=\pm1$. When the two-fold rotation symmetry is broken, the two Majorana modes from the two individual sectors may hybridize. Therefore, if $z^{(+)}_2(\bar\Gamma,\bar{M_{1,2,3}})=z^{(-)}_2(\bar\Gamma,\bar{M_{1,2,3}})$, there is no Majorana mode after adding the mass; and if $z^{(+)}_2(\bar\Gamma,\bar{M_{1,2,3}})\neq{}z^{(-)}_2(\bar\Gamma,\bar{M_{1,2,3}})$, the total number of zero energy Majorana modes at a high-symmetry point projection is one (nonzero), and it remains at zero energy as one moves away from a high-symmetry point to a generic point $\bar\bk$ due to PHS and $C_2$, forming a flat band in the whole SBZ. Hence the constraint between $z_2(\bar\bk)$ (the $Z_2$ invariant for a generic line) and $z^{(\pm)}_2(\bar\Gamma,\bar{M_{1,2,3}})$ is found to be $z_2(\bar{\bk})=z^{(+1)}(\bar\Gamma,\bar{M_{1,2,3}})+z^{(-1)}(\bar\Gamma,\bar{M_{1,2,3}})$. 

Our second, more intricate, example concerns a Hamiltonian in class DIII with $C_{3,0}$-symmetry. For $HK$ or $H'K'$, the index is $Z^3$, one integer for each of the three eigenspaces of $C_3$. This integer index (for $HK$ or for $H'K'$) equals the number of Majorana modes with corresponding eigenvalue of $+i$ of the chiral operator minus the number of $-i$ Majorana modes in each sector at $\bar{K},\bar{K}'$ [see Fig.\ref{fig:SBZ}(f)]. Therefore, when $C_3$ is broken, the total number of surface Majorana modes is given by $z(\bar\bk)=z^{(-1)}(\bar{K},\bar{K'})+z^{(-\omega)}(\bar{K},\bar{K'})+z^{(-\bar\omega)}(\bar{K},\bar{K'})$. For the line that projects to $\bar\Gamma$, both TRS and PHS are separately preserved, so the number of $+i$ MZMs must equal the number of $-i$ Majorana modes, dictating that $z(\bar\bk)=0$.

In 3D superconductors with time-reversal symmetry, classes CI and DIII, there is a $Z$-classification without the existence of any additional symmetry. This $Z$-index cannot be uniquely determined by the set of the above indices given by high-symmetry lines. This can be understood easily from the perspective of surface states: the invariants defined on high-symmetry lines are only related to the number and chirality of Majorana cones at high symmetry points in the surface Brillouin zone, while there can be protected Majorana cones away from these high-symmetry points. From this perspective, it is natural that the total chirality (a topological index) can be determined by the invariants on high-symmetry lines up to a multiple of $n$, because any Majorana cone centered at a generic $\bk$ on the surface must coexist with the other $n-1$ Majorana cones of the same chirality due to the $C_n$ symmetry.

\subsection{Effective Surface Theory}

On the surface, topologically non-trivial superconductors have protected surface Majorana cones containing Majorana modes that have a specific chirality. As we have mentioned previously, these Majorana cones may or may not appear at high-symmetry points in the surface BZ, and yet in either case, the specific $k\cdot{p}$ form of each cone is subject to constraints placed by symmetries. In this subsection, we study the constraints on these surface Majorana cones placed by the presence of rotational symmetries. In general, for the rotation symmetry to be relevant, we require that the surface is invariant under $C_n$. In most cases, the generic form of a surface Majorana cones reads
\bea
h(\bq)=d_0(\bq)+\sum_{i=x,y,z}d_i(\bq)\sigma_i.
\eea
The symmetries place constraints on the form of $d_{0,x,y,z}(\bk)$ and it is our goal to find these constraints in the presence of various symmetries including $C_{n,m}$, TRS and SU(2). As before, we are interested in studying two separate situations: (i) generic $\bk_0$ within the surface BZ and (ii) $\bk_0$ at high-symmetry points within the surface BZ.

\subsubsection{Generic $\bk_0$}

If there is no symmetry other than $C_n$ and PHS, then there is no constraint on the form of $h(\bq)$ for $n\in{odd}$, and a Majorana cone is not protected. Yet for $n\in{even}$, $P'$ as defined in Eq.(\ref{eq:newsym}) is a symmetry, which leads to
\bea\label{eq:49}
KP'h(\bq)(KP')^{-1}=-h(\bq).
\eea
In the $k\cdot{p}$ basis, $P'$ is represented by
\bea\label{eq:50}
P'=I,
\eea
and
\bea\label{eq:51}
P'=i\sigma_y
\eea
corresponding to situations of ${P'}^2=\mp1$, respectively. Substituting Eq.(\ref{eq:50}) or Eq.(\ref{eq:51}) into Eq.(\ref{eq:49}), we have
\bea
\label{eq:53}d_0(\bq)=d_x(\bq)=d_z(\bq)&=&0,\\
\label{eq:52}d_0(\bq)&=&0,
\eea
respectively. Therefore, if $m\in{even}$, we have ${KP'}^2=-1$ and the constraint Eq.(\ref{eq:52}) makes the co-dimension of the theory $1$ and, hence, fully gapped generically. On the other hand, if $m\in{odd}$, we have ${KP'}^2=1$, and the constraints Eq.(\ref{eq:53}) makes the co-dimension $-1$, which means the that theory is gapless along a certain direction and the Majorana modes could appear as nodal rings in the surface BZ.

Beyond this, we consider the addition of TRS in an effort to understand the manner in which the constraints change. The symmetries, in this case, of $h(\bq)$ are $KP'$ and $KT'$ if $n\in{even}$ or only the chiral symmetry, $S'=KP'*KT'$ if $n\in{odd}$. For $n\in{odd}$, $h(\bq)$ for a generic point $\bq$ in the surface BZ belongs to class AIII. For the symmetry representation, we choose $S=\sigma_z$, which leads to the constraint $d_0=d_z=0$. For $n\in{even}$, using Eqs.(\ref{eq:newsym}), we know that: (i) if $m\in{even}$, then $h(\bq)$ belongs to class CI, where $P'=(i\sigma_y)$ and $T'=I$ and (ii) if $m\in{odd}$, $h(\bq)$ belongs to class BDI, where $P'=\sigma_x$ and $T'=I$. Then by using
\bea
T'h(\bq){T'}^{-1}&=&h^\ast(\bq),\\
\nonumber
P'h(\bq){P'}^{-1}&=&-h^\ast(\bq),
\eea
we find the constraints that for $m\in{even}$, $d_0=d_y=0$ and for $m\in{odd}$, then $d_0=d_x=d_y=0$. When two out of four $d_i$'s are required to vanish, then the resultant co-dimension is zero, and a Majorana cone, if exists, is locally stable within the surface BZ. Furthermore, when three out of four $d_i$'s are required to vanish by the symmetry constraints, then the co-dimension is $-1$ and there is a nodal (Majorana) line in the surface bands. In all the discussion above, we have implicitly assumed that the vertical line which projects to a generic point is always trivial. In the case where a generic vertical line is non-trivial, there will be Majorana flat bands on the surface. By following this identical methodology, one can sweep out each of the separate permutations of symmetry constraints to derive effective theories around a generic point in the surface BZ, the results of which we summarize in Table \ref{tab:generic}.

\begin{table*}[htp]
\begin{tabular}{|c|c|c|c|c|c|}
\hline
Class & $KP'$ & $KT'$ & $S$ & $k\cdot{p}$ & Node type\\
\hline
C/D, $n\in{odd}$ & N/A & N/A & N/A & $d_x\sigma_x+d_y\sigma_y+d_z\sigma_z$ & Gapped\\
\hline
C/D, $(n\in{even},m\in{even})$ & $Ki\sigma_y$ & N/A & N/A & $d_x\sigma_x+d_y\sigma_y+d_z\sigma_z$ & Gapped\\
\hline
C/D, $(n\in{even},m\in{odd})$ & $K$ & N/A & N/A & $d_y\sigma_y$ & Nodal line\\
\hline
CI/DIII, $n\in{odd}$ & N/A & N/A & $\sigma_z$ & $d_x\sigma_x+d_y\sigma_y$ & Nodal point\\
\hline
CI/DIII, $(n\in{even},m\in{even})$ & $Ki\sigma_y$ & $K$ & $\sigma_y$ & $d_x\sigma_x$ & Nodal line\\
\hline
CI/DIII, $(n\in{even},m\in{odd})$ & $K$ & $K\sigma_z$ & $\sigma_z$ & $d_y\sigma_y$ & Nodal line\\
\hline
\end{tabular}
\caption{The effective surface theories in the vicinity of a generic point in the surface BZ for $3$D gapped superconductors constrained by the presence of $C_{n,m}$ symmetries and in the presence of particle-hole symmetry, time-reversal symmetry or $S$, the combination of particle-hole symmetry and time-reversal symmetry.}\label{tab:generic}
\end{table*}%

\subsubsection{$\bk_0$ at high-symmetry points}

Additional constraints on the effective $k\cdot{p}$ theory apply if $\bk_0$ is invariant under rotation, in other words $\bk_0$ is at a high-symmetry point within the surface BZ. As we have noted earlier, these high-symmetry points within the surface BZ may be either a two-, three-, four- or six-fold invariant points. While our goal is to provide a complete classification for the effective surface theories of gapped $3$D superconductors when the selected line within the surface BZ lies at a high-symmetry point, in what follows we address three specific situations in detail  for $(n,m)=(2,0)$, $(n,m)=(2,1)$ and $(n,m)=(4,2)$. We provide the complete results for arbitrary $(n,m)$ listed in Table \ref{tab:n=2},  \ref{tab:n=3}, \ref{tab:n=4}, \ref{tab:n=6} for $C_2$, $C_3$, $C_4$, and $C_6$ respectively.

\begin{table*}[htp]
\begin{tabular}{|c|c|c|c|c|c|}
\hline
Class & $KP$ & $KT$ & $C_{n,m}$ & $k\cdot{p}$ & Node type\\
\hline
C & $K(i\sigma_y)$ & N/A & $C_{2,0}=\pm{I}$ & $m_1\sigma_x+m_2\sigma_y+m_3\sigma_z$ & Gapped\\
\hline
C & $K(i\sigma_y)$ & N/A & $C_{2,1}=i\sigma_z$ & $m\sigma_z$ & Gapped\\
\hline
D & $K\sigma_x$ & N/A & $C_{2,0}=i\sigma_z$ & $m\sigma_z$ & Gapped\\
\hline
D & $K\sigma_x$ & N/A & $C_{2,1}=\pm1$ & $m\sigma_z$ & Gapped\\
\hline
D & $K\sigma_x$ & N/A & $C_{2,1}=\sigma_x$ & $(vk_x+wk_y)\sigma_y$ & Nodal line\\
\hline
CI & $K(i\sigma_y)$ & $K\sigma_x$ & $C_{2,0}=\pm{I}$ & $m_1\sigma_x+m_2\sigma_y$ & Gapped\\
\hline
CI & $K(i\sigma_y)$ & $K\sigma_x$ & $C_{2,1}=i\sigma_x$ & $m\sigma_x$ & Gapped\\
\hline
DIII & $K\sigma_x$ & $K(i\sigma_y)$ & $C_{2,0}=i\sigma_z$ & $A_{ij}k_i\sigma_j$ & Linear\\
\hline
DIII & $K\sigma_x$ & $K(i\sigma_y)$ & $C_{2,1}=\sigma_z$ & $A_{ij}k_i\sigma_j$ & Linear\\
\hline
\end{tabular}
\caption{The effective theories on the surface in the vicinity of a $C_2$-invariant point in the surface BZ for $3$D gapped superconductors constrained by the presence of $C_{n,m}$ symmetries and in the presence of particle-hole symmetry, time-reversal symmetry or $S$, the combination of particle-hole symmetry and time-reversal symmetry. An entry of 'N/A' indicates that the particular constraints considered in a particular tabular entry are not applicable.}\label{tab:n=2}
\end{table*}%

For $(n,m)=(2,0)$, there are four high-symmetry points in the surface BZ: $\bar\Gamma$ and $\bar{M_{1,2,3}}$, each of which is $C_2$-invariant. Therefore the analysis is the same for all four high-symmetry points. In class C, we utilize $KP^2=-1$ and $C_{2,0}^2=1$, in which case we have $P=(i\sigma_y)$ and $C_{2,0}=\sigma_0$. This shows that a mass term $m\sigma_{x,y,z}$ can be added to the surface theory so that it is gapped. Considering a class D system, using $P^2=1$ and $C_{2,0}^2=-1$, we have $P=\sigma_x$ and $C_2=i\sigma_z$, therefore, in this case, we may also have a mass term $m\sigma_z$ in the theory, effectively gapping the system again. In class CI, in addition to $C_{2,0}$ and $KP$, represented by the same matrices as in class C, we have $T=I$, so once again mass terms such as $m\sigma_{x,z}$ can be added, and the theory is gapped. However, in class DIII, we have $T=(i\sigma_y)$ and the resultant theory is massless with the lowest order expansion being $h(k_x,k_y)=A_{ij}k_i\sigma_j$ where $i,j=x,y$, typical of a Majorana cone.

\begin{table*}[htp]
\begin{tabular}{|c|c|c|c|c|c|}
\hline
Class & $KP$ & $KT$ & $C_{n,m}$ & $k\cdot{p}$ & Node type\\
\hline
C & $K(i\sigma_y)$ & N/A & $C_{3,0}=I$ & $m_1\sigma_x+m_2\sigma_y+m_3\sigma_z$ & Gapped\\
\hline
C & $K(i\sigma_y)$ & N/A & $C_{3,0}=\exp(i\frac{2\pi}{3}\sigma_z)$ & $m\sigma_z$ & Gapped\\
\hline
C & $K(i\sigma_y)$ & N/A & $C_{3,1},C_{3,2}=-I$ & $m_1\sigma_x+m_2\sigma_y+m_3\sigma_z$ & Gapped\\
\hline
C & $K(i\sigma_y)$ & N/A & $C_{3,1},C_{3,2}=-\exp(i\frac{2\pi}{3}\sigma_z)$ & $m\sigma_z$ & Gapped\\
\hline
D & $K\sigma_x$ & N/A & $C_{3,0}=-I$ & $m\sigma_z$ & Gapped\\
\hline
D & $K\sigma_x$ & N/A & $C_{3,0}=-\exp(i\frac{2\pi}{3}\sigma_z)$ & $m\sigma_z$ & Gapped\\
\hline
D & $K\sigma_x$ & N/A & $C_{3,1},C_{3,2}=I$ & $m\sigma_z$ & Gapped\\
\hline
D & $K\sigma_x$ & N/A & $C_{3,1},C_{3,2}=\exp(i\frac{2\pi}{3}\sigma_z)$ & $m\sigma_z$ & Gapped\\
\hline
CI & $K(i\sigma_y)$ & $K\sigma_x$ & $C_{3,0}={I}$ & $m_1\sigma_x+m_2\sigma_y$ & Gapped\\
\hline
CI & $K(i\sigma_y)$ & $K\sigma_x$ & $C_{3,0}=\exp(i\frac{2\pi}{3}\sigma_z)$ & $ck_+^2\sigma_-+h.c.$ & Quadratic\\
\hline
DIII & $K\sigma_x$ & $K(i\sigma_y)$ & $C_{3,0}=-I$ & $c_1k_+^3\sigma_++c_2k_-^3\sigma_-+h.c.$ & Cubic\\
\hline
DIII & $K\sigma_x$ & $K(i\sigma_y)$ & $C_{3,0}=-\exp(i\frac{2\pi}{3}\sigma_z)$ & $ck_+\sigma_-+h.c.$ & Linear\\
\hline
\end{tabular}
\caption{The effective theories on the surface in the vicinity of a $C_3$-invariant point within the surface BZ for $3$D gapped superconductors constrained by the presence of $C_{n,m}$ symmetries and in the presence of particle-hole symmetry, time-reversal symmetry or $S$, the combination of particle-hole symmetry and time-reversal symmetry. An entry of 'N/A' indicates that the particular constraints considered in a particular tabular entry are not applicable.}\label{tab:n=3}
\end{table*}%

We proceed to discuss the case in which $(n,m)=(2,1)$. Considering class C, we use the fact that $KP^2=C_{2,1}=-1$ in conjunction with $P=(i\sigma_y)$ and $C_{2,1}=i\sigma_z$, thereby allowing a mass term $m\sigma_z$ that results in the formation of a gap in the surface BZ. Moving to class D, we take advantage of $KP^2=C_{2,1}^2=1$, where we have $P=\sigma_x$ and $C_{2,1}=I$, or $C_{2,1}=\sigma_x$, which correspond to the cases where the two bands have the same and the opposite $C_{2,1}$ eigenvalues respectively. With the case of $C_{2,1}=I$, mass term $m\sigma_z$ may be added, however, for $C_{2,1}=\sigma_x$, mass terms are disallowed and the resultant lowest order surface theory is a linear term $(vk_x+wk_y)\sigma_y$, which indicates a nodal line in the SBZ passing through the high-symmetry point. This is consistent with the previous result that for $(n\in{even},m\in{odd})$, the effective theory around a generic point exhibits nodal lines in class C. With class CI, we have: $P=i\sigma_y$, $T=\sigma_x$ and $C_{2,1}=i\sigma_x$ where we have used $[KP,C_{2,1}]=\{KT,C_{2,1}\}=0$. In this case, a mass term $m_1\sigma_x+m_2\sigma_y$ can be added that naturally gaps the effective surface theory. Lastly, in class DIII, we have $P=\sigma_x$, $T=K=i\sigma_y$ and $C_{2,1}=\sigma_z$. The mass terms in class DIII are disallowed and the lowest order terms allowable in the surface theory are in the form $\sum_{ij}A_{ij}k_i\sigma_j$, which represents a Majorana cone.

\begin{table*}[htp]
\begin{tabular}{|c|c|c|c|c|c|}
\hline
Class & $KP$ & $KT$ & $C_{n,m}$ & $k\cdot{p}$ & Node type\\
\hline
C & $K(i\sigma_y)$ & N/A & $C_{4,0},C_{4,2}=\pm{I}$ & $m_1\sigma_x+m_2\sigma_y+m_3\sigma_z$ & Gapped\\
\hline
C & $K(i\sigma_y)$ & N/A & $C_{4,0},C_{4,2}=i\sigma_z$ & $m\sigma_z$ & Gapped\\
\hline
C & $K(i\sigma_y)$ & N/A & $C_{4,1},C_{4,3}=\pm\exp(i\frac{\pi}{4}\sigma_z)$ & $m\sigma_z$ & Gapped\\
\hline
D & $K\sigma_x$ & N/A & $C_{4,0},C_{4,2}=\pm\exp(i\frac{\pi}{4}\sigma_z)$ & $m\sigma_z$ & Gapped\\
\hline
D & $K\sigma_x$ & N/A & $C_{4,1},C_{4,3}=\pm1$ & $m\sigma_z$ & Gapped\\
\hline
D & $K\sigma_x$ & N/A & $C_{4,1},C_{4,3}=\sigma_x$ & $[r_1(k_x^2-k_y^2)+r_2k_xk_y]\sigma_z$ & Nodal line\\
\hline
D & $K\sigma_x$ & N/A & $C_{4,1}=i\sigma_z$ & $m\sigma_z$ & Gapped\\
\hline
CI & $K(i\sigma_y)$ & $K\sigma_x$ & $C_{4,0}=\pm{I}$ & $m_1\sigma_x+m_2\sigma_y$ & Gapped\\
\hline
CI & $K(i\sigma_y)$ & $K\sigma_x$ & $C_{4,0}=i\sigma_z$ & $c_1k_+^2\sigma_++c_2k_-^2\sigma_-+h.c.$ & Quadratic\\
\hline
CI & $K(i\sigma_y)$ & $K\sigma_x$ & $C_{4,2}=\sigma_z$ & $c_1k_+^2\sigma_++c_2k_-^2\sigma_-+h.c.$ & Quadratic\\
\hline
CI & $K(i\sigma_y)$ & $K\sigma_x$ & $C_{4,2}=i\sigma_x$ & $m\sigma_x$ & Gapped\\
\hline
DIII & $K\sigma_x$ & $K(i\sigma_y)$ & $C_{4,0}=\pm\exp(i\frac{\pi}{4}\sigma_z)$ & $ck_+\sigma_-+h.c.$ & Linear\\
\hline
DIII & $K\sigma_0\otimes\sigma_x$ & $K(i\sigma_y)\otimes\sigma_x$ & $C_{4,2}=\sigma_z\otimes\exp(i\frac{\pi}{4}\sigma_z)$ & $m\sigma_z\otimes\sigma_z$ & Gapped\\
\hline
\end{tabular}
\caption{The effective theories on the surface in the vicinity of a $C_4$-invariant point within the surface BZ for $3$D gapped superconductors constrained by the presence of $C_{n,m}$ symmetries and in the presence of particle-hole symmetry, time-reversal symmetry or $S$, the combination of particle-hole symmetry and time-reversal symmetry. Within the table an entry of 'N/A' indicates that the particular constraints considered in a particular tabular entry are not applicable.}\label{tab:n=4}
\end{table*}%

We note that there are several cases where a simple two-band model cannot adequately describe the symmetry groups. Put another way, this indicates that the symmetry group does not have any 2D irreducible representation. These cases are: (i) a class DIII Hamiltonian with $(n,m)=(4,2)$ (ii) a class CI Hamiltonian with $(n,m)=(6,3)$ and (iii) a class CI Hamiltonian with $(n,m)=(6,3)$, where in each case the smallest representation is four-dimensional. As an example, let us discuss class DIII with $(n,m)=(4,2)$; the other cases can be similarly discussed. Suppose we have one state in the $C_{4,2}=e^{i\pi/4}$-sector, then TRS takes it to the $-e^{-i\pi/4}$-sector [using $\{T,C_{4,2}\}=0$ from Eq.(\ref{eq:Cnm})], so we have two states with two different $C_{4,2}$-eigenvalues. Yet PHS will send these two states to another two states, in the $C_{4,2}=e^{-i\pi/4}$-sector and $-e^{i\pi/4}$-sector. Therefore, there must be at least four states, one in each sector, to realize the full symmetry group. We can choose the symmetries to be representsd by: $P=\sigma_0\otimes\sigma_x$, $T=(i\sigma_y)\otimes\sigma_x$ and $C_{4,2}=\sigma_x\otimes\exp(i\pi\sigma_z/4)$ and notice that a mass term $m\sigma_z\otimes\sigma_z$ is allowed, thereby rendering the effective surface theory gapped. Here the fact that the spectrum can be fully gapped and that the irreducible (projective) representation is at least four-dimensional do not contradict, as they would in a system without particle-hole symmetry (non-BdG Hamiltonian). In the latter case, a four-dimensional (or any higher than one-dimensional) representation implies that the single particle spectrum must be gapless. The distinction is because that for a BdG Hamiltonian, the operator $KP$ anti-commutes, rather than commutes with the Hamiltonian, and is hence not a real symmetry.

\begin{table*}[htp]
\begin{tabular}{|c|c|c|c|c|c|}
\hline
Class & $KP$ & $KT$ & $C_{n,m}$ & $k\cdot{p}$ & Node type\\
\hline
C & $K(i\sigma_y)$ & N/A & $C_{6,0},C_{6,2},C_{6,4}=\pm{I}$ & $m_1\sigma_x+m_2\sigma_y+m_3\sigma_z$ & Gapped\\
\hline
C & $K(i\sigma_y)$ & N/A & $C_{6,0},C_{6,2},C_{6,4}=\pm\exp(i\frac{\pi}{3}\sigma_z)$ & $m\sigma_z$ & Gapped\\
\hline
C & $K(i\sigma_y)$ & N/A & $C_{6,1},C_{6,3},C_{6,5}=\pm\exp(i\frac{\pi}{6}\sigma_z)$ & $m\sigma_z$ & Gapped\\
\hline
C & $K(i\sigma_y)$ & N/A & $C_{6,1},C_{6,3},C_{6,5}=i\sigma_z$ & $m\sigma_z$ & Gapped\\
\hline
D & $K\sigma_x$ & N/A & $C_{6,0},C_{6,2},C_{6,4}=\pm\exp(i\frac{\pi}{6}\sigma_z)$ & $m\sigma_z$ & Gapped\\
\hline
D & $K\sigma_x$ & N/A & $C_{6,0},C_{6,2},C_{6,4}=i\sigma_z$ & $m\sigma_z$ & Gapped\\
\hline
D & $K\sigma_x$ & N/A & $C_{6,1},C_{6,3},C_{6,5}=\pm{I}$ & $m\sigma_z$ & Gapped\\
\hline
D & $K\sigma_x$ & N/A & $C_{6,1},C_{6,3},C_{6,5}=\sigma_x$ & $ck_+^3\sigma_y+h.c.$ & Nodal line\\
\hline
D & $K\sigma_x$ & N/A & $C_{6,1},C_{6,3},C_{6,5}=\pm\exp(i\frac{\pi}{3}\sigma_z)$ & $m\sigma_z$ & Gapped\\
\hline
CI & $K(i\sigma_y)$ & $K\sigma_x$ & $C_{6,0}=\pm{I}$ & $m_1\sigma_x+m_2\sigma_y$ & Gapped\\
\hline
CI & $K(i\sigma_y)$ & $K\sigma_x$ & $C_{6,0}=\pm\exp(i\frac{\pi}{3}\sigma_z)$ & $ck_+^2\sigma_-+h.c.$ & Quadratic\\
\hline
CI & $K\sigma_z\otimes(i\sigma_y)$ & $K\sigma_x\otimes\sigma_x$ & $C_{6,3}=\sigma_z\otimes\exp(i\frac{\pi}{6}\sigma_z)$ & $m\sigma_0\otimes\sigma_z$ & Gapped\\
\hline
CI & $K(i\sigma_y)$ & $K\sigma_x$ & $C_{6,3}=i\sigma_x$ & $m\sigma_x$ & Gapped\\
\hline
DIII & $K\sigma_x$ & $K(i\sigma_y)$ & $C_{6,0}=\pm\exp(i\frac{\pi}{6}\sigma_z)$ & $ck_+\sigma_-+h.c.$ & Linear\\
\hline
DIII & $K\sigma_x$ & $K(i\sigma_y)$ & $C_{6,0}=i\sigma_z$ & $c_1k^3_+\sigma_++c_2k^3_-\sigma_-+h.c.$ & Cubic\\
\hline
DIII & $K\sigma_x$ & $K(i\sigma_y)$ & $C_{6,3}=\sigma_x$ & $ck_+^3\sigma_y+h.c.$ & Nodal line\\
\hline
DIII & $K\sigma_0\otimes\sigma_x$ & $K(i\sigma_y)\otimes\sigma_x$ & $C_{6,3}=\sigma_z\otimes\exp(i\frac{\pi}{3}\sigma_z)$ & $m\sigma_z\otimes\sigma_z$ & Gapped\\
\hline
\end{tabular}
\caption{The effective theories on the surface in the vicinity of a $C_6$-invariant point within the surface BZ for $3$D gapped superconductors constrained by the presence of $C_{n,m}$ symmetries and in the presence of particle-hole symmetry, time-reversal symmetry or $S$, the combination of particle-hole symmetry and time-reversal symmetry. Within the table an entry of 'N/A' indicates that the particular constraints considered in a particular tabular entry are not applicable.}\label{tab:n=6}
\end{table*}%

\section{Conclusion}\label{sec:conclusion}
In conclusion, we have examined the topological properties of the Bogoliubov - de Gennes Hamiltonians from the Altland-Zimbauer classification scheme corresponding to gapped topological superconductors in one, two and three spatial dimensions in the presence of rotational symmetry. In $1$D, we complete the classification by block-diagonalizing the Hamiltonian into sectors labeled by rotation eigenvalues, and finding the topological index, $Z$ or $Z_2$ or trivial, for each sector. The role played by the angular momentum of the Cooper pairs is emphasized: it generally leads to a nontrivial projective representation of rotation symmetry, where anomalous commutation relations include $C_n^2\neq(-1)^F$ and the anti-commutation between the time-reversal and the rotation. In $2$D, we explicitly proved the relation between the Chern number and the rotation eigenvalues at high-symmetry points, and showed that in the weak-coupling limit, the contribution breaks down to two parts. One part is from the normal state band structure the other part is from the angular momentum of the Cooper pairs. In $3$D, we found the bulk topological invariants for all rotational symmetries, linearly and projectively represented, defined on high-symmetry lines and generic lines; we correspondingly found the relevant surface theories for both generic and high-symmetry points in the surface Brillouin zone.

\begin{acknowledgements}
MJG would like to thank Dan Arovas for enlightening conversations. CF was supported by Project 11674370 by NSFC and the National Key Research and Development Program under grant number 2016YFA0302400 and 2016YFA0300600. CF, and MJG were supported by ONR - N0014-11-1-0123 and MJG was supported by NSF-CAREER EECS-1351871. BAB was supported by De- partment of Energy de-sc0016239, NSF EAGER Award NOA - AWD1004957, Simons Investigator Award, ONR - N00014-14-1-0330, ARO MURI W911NF-12-1-0461, NSF-MRSEC DMR-1420541, Packard Foundation and Schmidt Fund for Innovative Research.
\end{acknowledgements}

\begin{widetext}

\appendix

\section{Linear and projective representations of the group generated by $P$ and $C_n$}
\label{app:PC}
A linear representation of the group generated by $P$ and $C_n$ must satisfy
\bea\label{eq:linearRP}
D(P)D^\ast(P)&=&1,\\
\nonumber
D^n(C_n)&=&(-1)^F,\\
\nonumber
D(C_n)D(P)&=&D(P)D^\ast(C_n).
\eea
All representations that do not obey these relations, but obey them up to some phase factor, are called projective representations. However, there is a class of projective representations that are trivial, as they can be transformed back into a linear representation by multiplying each element some phase factor. If a projective representation cannot be brought back to a linear representation, it is called a nontrivial projective representation.

We prove by contradiction that if a projective representation satisfies
\bea
D(C_n)D(P)=D(P)D^\ast(C_n),\\
\nonumber
D^n(C_n)=-(-1)^F,
\eea
then it must be nontrivial.

Suppose it is trivial, then we can redefine the generators as
\bea\label{eq:projRP}
D(P)\rightarrow{D}(P)e^{i\omega(P)},\\
\nonumber
D(C_n)\rightarrow{D}(C_n)e^{i\omega(C_n)},
\eea
so that Eq.(\ref{eq:linearRP}) becomes
\bea
D(P)D^\ast(P)&=&1,\\
\nonumber
D^n(C_n)&=&e^{in\omega(C_n)}(-1)^F,\\
\nonumber
D(C_n)D(P)e^{i\omega(C_n)}&=&D(P)D^\ast(C_n)e^{-i\omega(C_n)}.
\eea
In order to satisfy Eq.(\ref{eq:projRP}), one requires
\bea
\omega(C_n)=m\pi/n,
\omega(C_n)=0,\pi,
\eea
where $m\in{odd}$. It is obvious that the two equations are contradictory. Therefore, a representation satisfying Eq.(\ref{eq:projRP}) is a nontrivial projective representation.

\section{Linear and projective representation of the group generated by $P$, $T$, and $C_n$}
\label{app:PCT}
A linear representation of the group should satisfy, besides the equations in Eq.(\ref{eq:linearRP})
\bea
D(P)D^\ast(T)&=&D(T)D^\ast{}(P),\\
\nonumber
D(T)D^\ast(C_n)&=&D(C_n)D(T),\\
\nonumber
D(T)D^\ast(T)&=&(-1)^F.
\eea
We prove that when $n\in{even}$, representations that satisfy
\bea\label{eq:projRP2}
D(T)D^\ast(D_n)&=&-D(C_n)D(T),\\
\nonumber
D(P)D^\ast(C_n)&=&D(C_n)D(P).
\eea
must be nontrivial.

Again we first assume that it is trivial, so that one can find $\omega(T)$ and $\omega(C_n)$ such that Eq.(\ref{eq:projRP2}) is satisfied. Multiply $D(T)$ and $D(C_n)$ with phase factors $e^{i\omega(T)}$ and $e^{i\omega(C_n)}$ so that they satisfy
\bea
D(T)D^\ast(C_n)e^{-i\omega(C_n)}&=&D(C_n)D(T)e^{i\omega(C_n)},\\
\nonumber
D(P)D^\ast(C_n)e^{-i\omega(C_n)}&=&D(C_n)D(P)e^{i\omega(C_n)}.
\eea
In order to satisfy Eq.(\ref{eq:projRP2}), one requires, respectively
\bea
\omega(C_n)&=&\pm\pi,\\
\nonumber
\omega(C_n)&=&0, \pi
\eea
which obviously contradict each other. Therefore, a representation satisfying Eq.(\ref{eq:projRP2}) is indeed a nontrivial projective representation.

\section{General Proof of Chern Number for $C_{\infty}$ Rotational Invariance}
\label{app:cproof}

Here we assume that the wavefunction, or more precisely, the projector onto the occupied space at $\bk=\infty$ is well defined, which is denoted by $P(\infty)$. This ensures a closed manifold, which is necessary for a well regularized Chern number calculation.

Consider a loop that consists of three parts: (i) a straight line from $\bk=0$ to a very large $\bk_1$, the azimuthal angle of which is zero, (ii) an arc going counterclockwise by $\delta\theta$ from $\bk_1$ to $\bk_2$and (iii) a straight line going from $\bk_2$ back to $\bk=0$ (see Fig.\ref{fig:Appn}). 
\begin{figure}
\includegraphics[width=8cm]{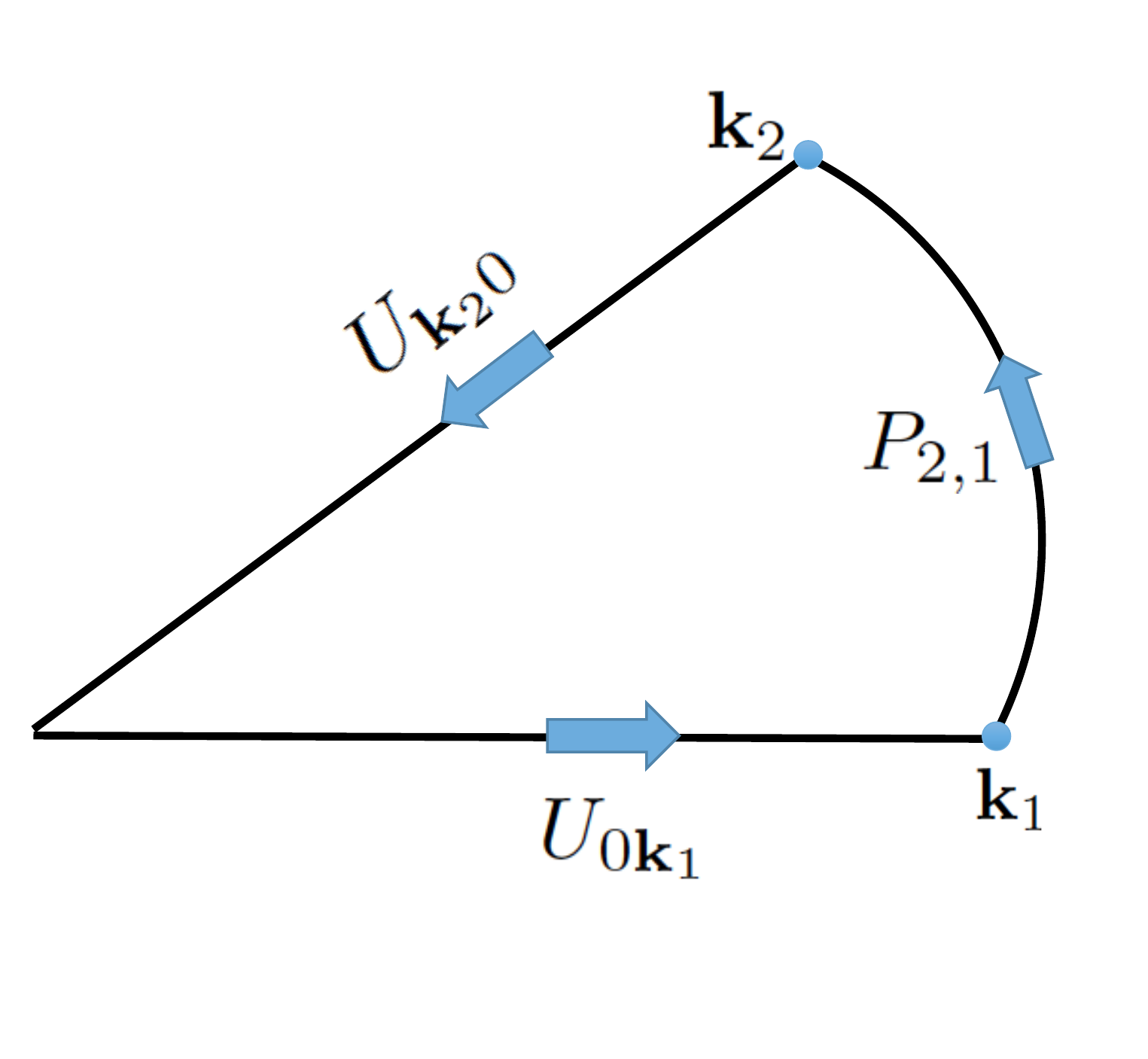}
\caption{\label{fig:Appn}A closed path that encircles a sector of angle $\theta$.}
\end{figure}
Calling them segments 1, 2, 3 and dividing each segment into $N_{1,2,3}$ smaller segments, the Wilson loop of the above loop is
\bea
W&=&U_1U_2U_3,
\eea
where
\bea
U_1&=&\prod_{i=1,...,N_1}P_{1,i},\\
U_2&=&\prod_{i=1,...,N_2}P_{2,i},\\
U_3&=&\prod_{i=1,...,N_3}P_{3,i},
\eea
where $P_{n,i}$ is the projection operator onto the occupied states at the $i$-th point of the $n$-th segment.
Since $P(\infty)$ is well defined, we have that when $|\bk_1|=|\bk_2|\rightarrow\infty$, we have $P_{2,i}=P_{2,j}$, or
\bea
U_2&=&P_{2,1}.
\eea
Therefore
\bea
\lim_{|\bk_1|\rightarrow\infty}W=U_{0\bk_1}U_{\bk_20}.
\eea
Then we notice that
\bea
U_{0\bk_2}=R(\delta\theta)U_{0\bk_1}R(-\delta\theta),
\eea
where $R(\theta)$ is the rotation operator through $\theta$,
and that
\bea
\lim_{|\bk_1|\rightarrow\infty}R(\delta\theta)U_{0\bk_1}R(-\delta\theta)=e^{ij(0)\delta\theta}U_{0\bk_1}e^{-ij(\infty)\delta\theta}.
\eea
From all above equations and using $U_{0\bk_1}U_{\bk_10}=1$, we have
\bea
\lim_{|\bk_1|\rightarrow\infty}W=\exp[i(j(0)-j(\infty))\delta\theta].
\eea
According to the relation between the Berry curvature integral and the Wilson loop we know that
\bea
\int_0^\infty{k}dk\int_0^{\delta\theta}d\theta{F}(k,\theta)&=&2\pi{n}+(j(0)-j(\infty))\delta\theta\\
\nonumber&=&(j(0)-j(\infty))\delta\theta.
\eea
The last equality uses the fact that for very small $\delta\theta$, the integral must also be very small, so the integer part is zero. The $C_{\infty}$ symmetry, the Berry's curvature is also rotationally invariant, so
\bea
C=\frac{1}{2\pi}\int{dk^2}F(k,\theta)=j(0)-j(\infty).
\eea

\end{widetext}

\end{document}